\documentclass[12pt,preprint]{aastex}

\usepackage{emulateapj5}
\usepackage{epsfig}
\usepackage{apjfonts}

\makeatletter

\newenvironment{figurehere}
  {\def\@captype{figure}}
  {}
\makeatother

\input{psfig.sty}

\def\Halpha{H$\alpha$~}
\def\about{$\sim$}
\def\arcsec{$\,^{\prime\prime}$~}
\def\arcmin{$\,^\prime$~}

\def\erg/cm2sec{ergs~cm$^{-2}$~s$^{-1}$}  
\def\ergcm2{ergs~cm$^{-2}$}  
  
\def\mdot{$\dot{m}$~}  
\def\X{$\times$~}

\def\Lx{L$_x$~}

\def\pc3{pc$^{-3}$~}

\def\cm-3{cm{$^{-3}$}~} 
\def\km/s{km~s$^{-1}$~}
\def\Pdot{{$\dot{P}$}~}
\def\Edot{{$\dot{E}$}~}
\def\rcore{$r_c$~}
\def\X{$\times$}

\def\WAVDETECT{{\sc wavdetect}~}
\def\vdisp{$\sigma_{v0}$}

\def\apj{ApJ}
\def\aa{A\&A}

\newcommand{\lsim }{{\lower0.8ex\hbox{$\buildrel <\over\sim$}}}
\newcommand{\gsim }{{\lower0.8ex\hbox{$\buildrel >\over\sim$}}}

\newcommand{\Msun}{\ifmmode {M_{\odot}}\else${M_{\odot}}$\fi~}
\newcommand{\Rsun}{\ifmmode {R_{\odot}}\else${R_{\odot}}$\fi~}
\newcommand{\Lsun}{\ifmmode {L_{\odot}}\else${L_{\odot}}$\fi~}
\newcommand{\mv}{\ifmmode {m_{V}}\else${m_{V}}$\fi}
\newcommand{\Mv}{\ifmmode {M_{V}}\else${M_{V}}$\fi}
\newcommand{\lopt}{\ifmmode L_{opt} \else $~L_{opt}$\fi}
\newcommand{\loglopt}{\ifmmode{\rm log}~L_{opt} \else log$~L_{opt}$\fi}
\newcommand{\lx}{\ifmmode L_x \else $~L_x$\fi}
\newcommand{\loglx}{\ifmmode{\rm log}~L_x \else log$~L_x$\fi}
\newcommand{\cmsq}{\ifmmode{\rm ~cm^{-2}} \else cm$^{-2}$\fi}
\newcommand{\nh}{\ifmmode{\rm N_{H}} \else N$_{H}$\fi}
\newcommand{\fcgs}{\ifmmode {\rm erg~cm}^{-2}~{\rm s}^{-1}\else
erg~cm$^{-2}$~s$^{-1}$\fi} 
\newcommand{\lcgs}{\ifmmode erg~~s^{-1}\else erg~s$^{-1}$\fi}

\begin{document}

\title{Chandra Study of a Complete Sample of Millisecond Pulsars in
47~Tuc and NGC~6397}

\author{J.~E. Grindlay\altaffilmark{1}, F. Camilo\altaffilmark{2},
C.~O. Heinke\altaffilmark{1}, P.~D. Edmonds\altaffilmark{1}, H.
Cohn\altaffilmark{3} and P. Lugger\altaffilmark{3}}
\altaffiltext{1}{Harvard-Smithsonian Center for Astrophysics, 60 Garden
  Street, Cambridge, MA~02138; josh@cfa.harvard.edu;
  cheinke@cfa.harvard.edu; pedmonds@cfa.harvard.edu}
\altaffiltext{2}{Columbia Astrophysics Laboratory, Columbia University,
  550 West 120th Street, New York, NY~10027;
  fernando@astro.columbia.edu}
\altaffiltext{3}{Department of Astronomy, Indiana University,
  Bloomington, IN~47405; cohn@astro.indiana.edu;
  lugger@astro.indiana.edu}

\received{2001 November 30}
\accepted{}

\begin{abstract}
We report Chandra observations of the complete sample of millisecond 
pulsars (MSPs) with precise radio positions in the globular clusters 
47~Tuc (NGC~104) and NGC~6397. 
The x-ray luminosities and colors are derived or constrained and
compared to x-ray MSPs previously detected in the field as well as 
one previously detected in a globular cluster (M28). The 47~Tuc
MSPs are predominantly soft sources suggestive of thermal emission from
small (r$_x <$ 0.6km) 
polar caps on 
the neutron star rather than magnetospheric emission and are a
relatively homogeneous sample, with most x-ray luminosities in a
surprisingly narrow range (\Lx \about1--4 \X 10$^{30}$ \lcgs). We
use previously derived intrinsic \Pdot values and find a new relation 
between \Lx and spindown luminosity, \Edot: \Lx $\propto$ \Edot$^{\beta}$, 
with $\beta$ \about0.5$\pm$0.2 vs. \about1.0 for both pulsars and MSPs in the
field. Adding the single MSP in NGC~6397 
constrains $\beta = 0.5\pm0.15$. 
This  \Lx--\Edot relation and also the  \Lx/\Edot vs. 
spindown age are each similar to that found by Harding 
\& Muslimov (2002) for thermal emission from polar cap heating.  
However, the cluster MSPs are relatively 
longer-lived (in thermal x-rays) 
than either the models or field MSPs, which may have additional 
magnetospheric (non-thermal) components. 
We suggest the cluster MSPs may have altered surface magnetic field 
topology (e.g. multipole) or their neutron stars 
are more massive from repeated accretion episodes due to 
encounters and repeated exchange interactions. 
MSP binary companions on or just off the 
main sequence (e.g. NGC~6397) are likely to have 
been re-exchanged and  
might show anomalous \Pdot and \Edot values  
due to relaxation of misaligned core-crust spins. The radial distribution of 
\about40 soft Chandra sources in 47~Tuc is 
consistent with a $\sim1.4~\Msun$ component in a multi-mass 
King model and with the identified MSP sample. 
The implied total MSP
population in 47~Tuc with \Lx \gsim10$^{30}$ \lcgs\ is \about35-90,  
and can constrain the relative beaming in radio vs. soft x-rays. 
 NGC~6397 is relatively deficient in MSPs; its single
detected example may have been re-exchanged out of the cluster core.

\end{abstract}


\section{INTRODUCTION}
Millisecond pulsars (MSPs) are end-products of binary evolution.  In
the Galactic disk, they are the fossil markers of rare primordial
binaries in which neutron stars (NSs) are retained after birth in
systems with low mass companions, from which accretion and a low mass
x-ray binary (LMXB) phase transfers angular momentum to spin them up to
msec periods. The ``standard model'' of MSPs as the remnants of LMXBs
(reviewed by Bhattacharya \& van den Heuvel 1991) is supported by the
discovery of the msec x-ray pulsar XTE~J1808$-$369 (Wijnands \& van der
Klis 1998) as well as the accumulating evidence for msec pulsation
periods in a number of LMXBs from apparently stable frequencies in
their quasi-periodic oscillation power spectra. However formation and
evolution questions remain since no LMXB in quiescence (qLMXB) has yet
been found as a radio MSP, and the reduction in angular momentum
transfer efficiency imposed by likely advection dominated accretion
flow (ADAF)-type accretion in quiescence means that the MSPs predominantly
arise from a population of bright LMXB transients (Yi \& Grindlay 1998)
which may not be observed in numbers sufficient to support the disk MSP
population.  The recent radio detection of 20 MSPs in the globular
cluster 47~Tuc (NGC~104) by Camilo et al.~(2000), and the 
subsequent x-ray detection with Chandra 
(Grindlay et al.~2001a; hereafter GHE01a) of the 15 that were
precisely located with radio timing solutions (Freire et al. 2001a;
hereafter FCL01), have opened a new window for
the study of the nature and formation/evolution of MSPs. 

In globular clusters, in which MSPs are expected to show the same
factor of \about100--200 excess in their numbers (per unit stellar
mass) as do the LMXBs compared with the disk population (Clark 1975),
the MSPs trace more exotic binary evolution (and possibly stellar
evolution) paths than in the field.  Here the MSP progenitor binaries
can be formed readily from main sequence binaries by the exchange of a
NS into the binary (e.g.  Rasio, Pfahl, \& Rappaport 2000) or they may
also (in very high density cluster cores) form from direct 2-body tidal
capture (Mardling 1995) or 3-body encounters.  Clusters also allow the
exchange of the original mass donor with a replacement 
main sequence (or near turnoff, given mass segregation) 
star, for which the MSP discovered in NGC~6397 (D'Amico et al.~2001a;
D'Amico et al.~2001b; hereafter DPM01; Ferraro et al.~2001) and 
detected with Chandra (Grindlay et al.~2001b; hereafter
GHE01b) might be the first example. 

The 15 MSPs with precise locations (FCL01) include 47~Tuc-G and -I
separated by just $0\farcs12$. The sample is  now extended to 16 with
the timing position for 47~Tuc-S (Freire 2001), which places it just
$0\farcs7$ from 47~Tuc-F.  We have identified (GHE01a) probable x-ray
emission from the original 15 MSPs (-G and -I cannot be resolved, and
-C is a marginal identification at best, as described below), and we
report here the possible detection of 47~Tuc-S.  
In this paper, we provide the
details for the Chandra detections of this complete 
(for radio detection)
sample of MSPs and derive the x-ray properties of this first
significant sample of MSPs detected in a single globular.  We derive
the x-ray luminosities, spindown energy loss $\dot E$, and the
characteristic age ($P/2\dot P$) for these MSPs, using the 47~Tuc
distance, 4.5$\pm0.3$ kpc, derived 
from the recent compilation of Zoccali et al.~(2001).  This is the
first derivation of these properties for a significant sample of MSPs
at a common known distance.  
We also compare the 47~Tuc MSP results to
those of the single MSP detected in NGC~6397 (GHE01b) and previous
(ROSAT) results for MSPs in the field and one other globular (M28), as
summarized in the review 
by Becker \& Tr\"umper (1999; hereafter BT99).

\section{CHANDRA OBSERVATIONS AND ANALYSIS OF MSPs IN 47~TUC}
A 74ksec Chandra observation (72ksec total data obtained) was conducted
on 2000 March 16--17, in which at least 108 sources were detected in
the $2' \times 2\farcm5$ around the cluster center (GHE01a).  With the
ACIS-I detector used, the $0\farcs492$ pixel size and $0\farcs7$ (FWHM)
telescope-detector point spread function (psf) for sources within 
\about4\arcmin of on-axis combine to give source positions with
1$\sigma$ centroid uncertainties of \lsim$0\farcs3$ for sources
detected with \gsim90\% confidence (Jerius et al.~2000). Even a source
identified with just 1 count (cf. discussion below of MSP 47~Tuc-C) is
thus located with $\sim0\farcs5$ 
(1$\sigma$) 
centroiding error.  The 
uncertainties in the 
absolute positions of Chandra sources are dominated by aspect (guide star)
offsets (i.e. boresight offsets between the x-ray and optical axes) and
are currently estimated to be $\sim 0\farcs7$ (Aldcroft et al. 2000).

\vspace*{0.12in}
\begin{figurehere}
\hspace*{-0.1in}
\epsfig{file=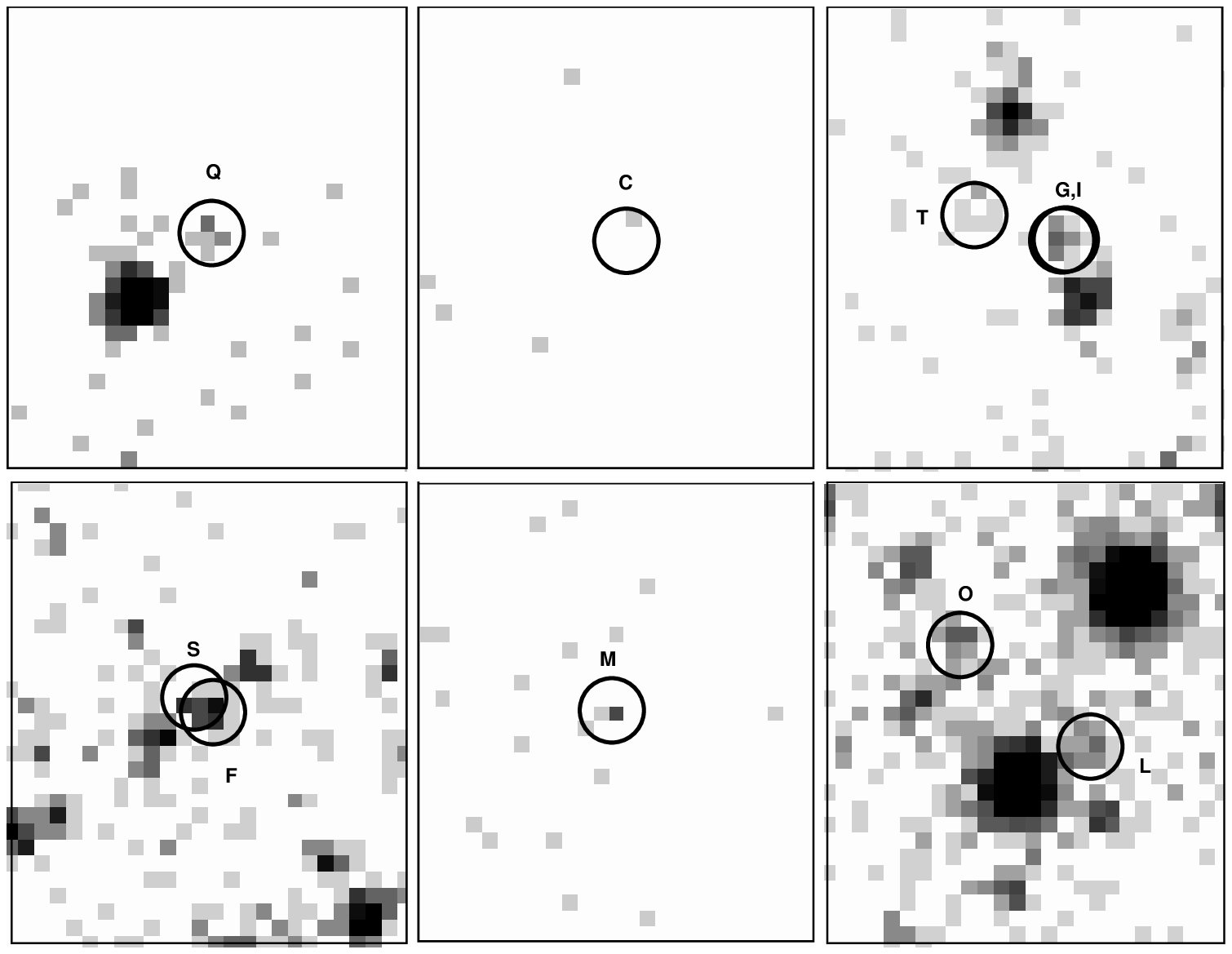,width=3.3in}
\caption{\label{fig:6msps} Images (0.5--4.5\,keV) of the MSPs in 47~Tuc
not detected (or confused) by \WAVDETECT (Q, C, T, G \& I, F \& S, and
L), showing probable emission from the MSP in each case, and the
weakest detection (M).  Pixel randomization has been removed from the
pipeline processing, and 1\arcsec circles are 
placed on the radio MSP locations.
These circles are used to extract the counts for each MSP, as they
contain $\sim95\%$ of the encircled energy for \lsim0.3\,keV sources, which 
describe all of the MSP spectra except for -J (cf. Fig. 2). }
\vspace*{0.14in}
\end{figurehere}

Given these relative and absolute errors for source positions, the
initial run of the CIAO\footnote{All CXC software is available at
http://asc.harvard.edu/ciao.} tool \WAVDETECT (Dobrzycki et al.~1999)
in our general study (GHE01a) of the sources near the center of 47~Tuc
detected 10 sources containing 11 of the MSPs with \about3--10\,mas
positions as given by FCL01.  A consistent offset of (RA, Dec) =
$(0\farcs38\pm0\farcs14, -0\farcs82\pm0\farcs11)$
was found for 6 relatively isolated MSPs
(47~Tuc-D, -E, -H, -J, -N, -U) detected with {\sc wavdetect}. Applying
this offset to correct the Chandra positions to the MSP reference
frame, we derived a small linear correction of the Chandra coordinates
in both RA and Dec to minimize the rms deviations 
(with RA, Dec values $0\farcs08, 0\farcs11$ for the positions 
given in Table 1)
with the radio positions.  
Details of this astrometry are given by Edmonds et
al.~(2001), who extended the astrometric solution to HST data to
identify the likely optical counterpart for 47~Tuc-U. For the 9
remaining sources of FCL01, including the unresolved pair -G and -I, we
used the precise astrometric solution to examine the Chandra images (in
each band) for evidence of these MSPs 
(cf. Fig.~\ref{fig:6msps}). 
For 3 of these (47~Tuc-L, -Q,
-T), \WAVDETECT failed to detect the source due to crowding with a
brighter neighbor, although inspection showed clumps of photons
(similar in color and flux to the detected MSPs) centered at the radio
position and consistent with the expected psf. For 47~Tuc-C only one
count is detected at the predicted MSP position, but the relatively
large exposure correction (factor of 2.5 [cf. Table~\ref{tab:results}],
due to its being near a detector gap) for this source is consistent
with it having x-ray flux within a factor of \about2 of the faintest
other MSP, 47~Tuc-M.  

Total counts were estimated for each MSP by extracting counts in each
band in a weighted (by the psf) 1\arcsec radius cell centered on the
MSP position and allowing for psf overlap contributions from
neighboring sources using their \WAVDETECT counts 
in each band and offset
positions.  The recently derived position for 47~Tuc-S (Freire 2001)
shows that the original \WAVDETECT flux (and thus derived x-ray colors
and L$_x$) for 47~Tuc-F (Table~1 of GHE01a) is contaminated by 47~Tuc-S,
separated by just $0\farcs7$. \WAVDETECT identified an elongated source
between the positions of 47~Tuc-F and 47~Tuc-S, which can be best
explained as the combined emission of the two sources.  We estimate the
counts for -F and -S by noting that the x-ray position is twice as
close to -F as -S, and thus assign -F two-thirds of the counts. This is
supported by visual inspection of the images in each band.  Detected
position offsets in the 0.5--4.5\,keV primary detection band for the 10
\WAVDETECT sources (12 MSPs) and counts in 3 bands (see below) for all
16 MSPs, are given in Table~\ref{tab:results}.

\subsection{X-ray Colors and Emission Models}
Our initial analysis of the MSP sample indicated the objects 
are
relatively soft sources compared with the probable cataclysmic
variables (CVs; GHE01a). In Table~\ref{tab:results} we give the detected
counts in 3 bands: softcts (0.2--1\,keV), mediumcts (1--2\,keV) and
hardcts (2--8\,keV) for each of the 14 resolved MSPs, with counts for
47~Tuc-G and -I (unresolved) estimated.  We use 3 bands rather than the
hardness ratio Xcolor = 2.5log(cts[0.5--1.5\,keV]/cts[1.5--6.0\,keV])
used in GHE01a since it allows an approximate spectral analysis for the 
MSPs.  With (limited) counts in 3 bands, we form the hardness ratios
HR1 = mediumcts/softcts and HR2 = hardcts/mediumcts and plot the MSPs,
with counting statistics errors, in the color-color diagram shown in
Figure~\ref{fig:cc}.  The MSP colors are clustered in a relatively
narrow range of HR1 and HR2; 47~Tuc-J is clearly (much) harder, as was
evident in the Xcolor distributions plotted in GHE01a.

For comparison with the data, we construct (using the PIMMS tool for
ACIS-I) values of HR1 and HR2 for 3 simple models
derived for the 47~Tuc absorption column ($N_H$ = 2.4 \X 10$^{20}$  
\cmsq; cf. GHE01a): 
thermal bremsstrahlung (TB), blackbody (BB) and power law (PL), with index
values (kT or photon index) given in the caption of
Figure~\ref{fig:cc}. It can be seen that the observed range of HR1 vs. HR2
is roughly consistent with TB spectra with kT \about 1\,keV, BB spectra
with kT \about 0.2--0.3\,keV or PL spectra with photon index \about3.
The track for BB models is most consistent with the data for all but
47~Tuc-J. The weighted mean colors are plotted for all the MSPs and for
all but 47~Tuc-J.  This is consistent with a single kT \about0.22\,keV
for all the MSPs (except -J). Thus we give in Table~\ref{tab:results}
approximate x-ray luminosity values for a BB spectrum with kT =
0.22\,keV. Apart from 47~Tuc-C, all other MSPs are detected with \Lx in
a surprisingly narrow range:  \about 1--4 \X 10$^{30}$ \lcgs, where the
\Lx values are again (as in GHE01a) quoted for the 0.5--2.5\,keV band to
facilitate comparison with ROSAT results for MSPs (BT99) and to
minimize uncertainties due to extrapolation of uncertain spectra over a
broad band in which counts are not actually detected.

\begin{figurehere}
\epsfig{file=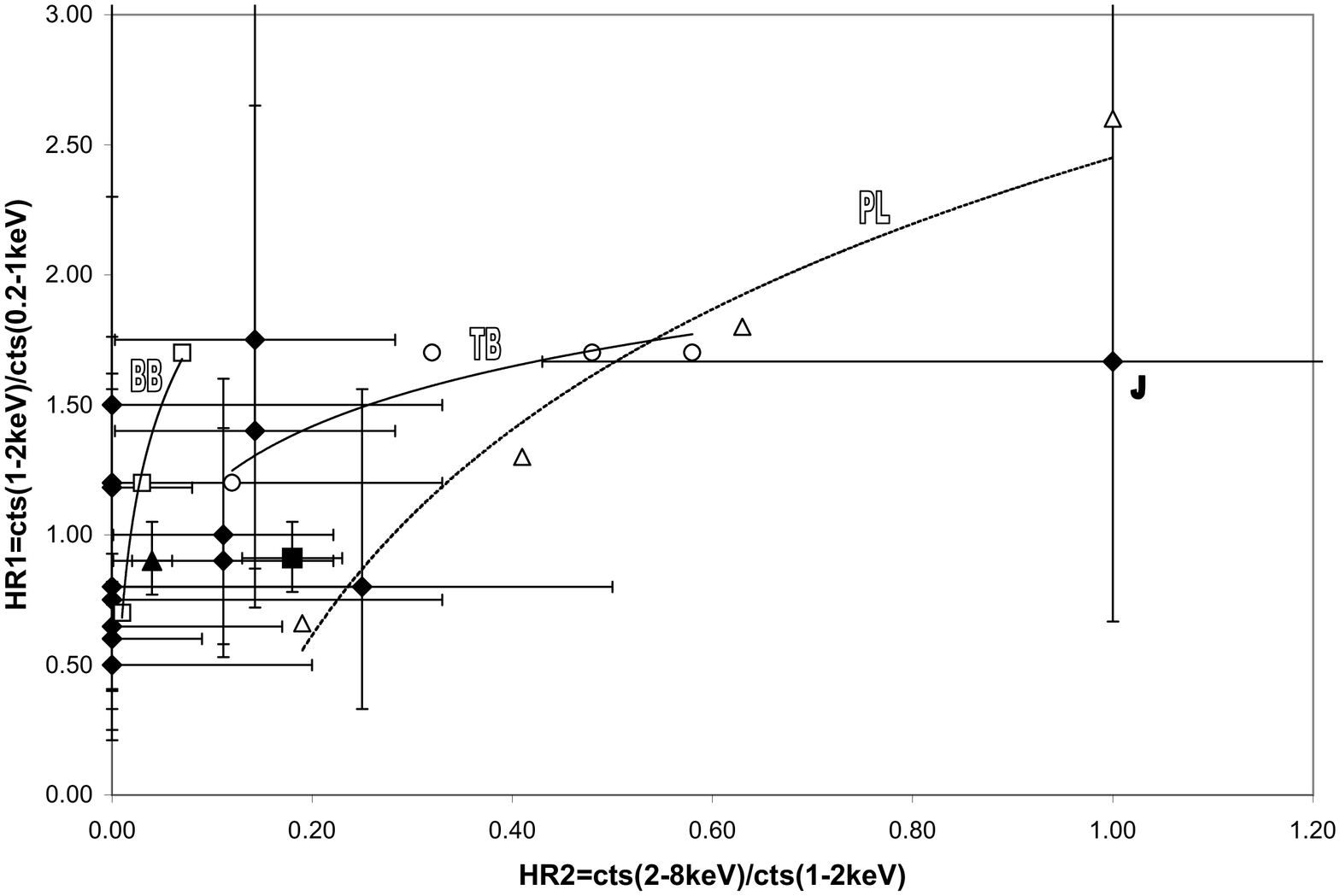,width=2.2in,angle=-90.}
\caption{\label{fig:cc} X-ray color-color diagram for all 
located MSPs (except 47~Tuc-C which is too faint, and -G which is 
combined with -I) in 47~Tuc 
(solid diamonds) vs. tracks for models with parameters changing from
lower left to upper right: blackbody (BB; open squares), with kT = 0.2,
0.25, 0.3\,keV; power law (PL; open triangles), with photon index
$\alpha$ = 3, 2, 1.5, 1; and thermal bremsstrahlung (TB; open circles),
with kT = 1, 2, 3, 6\,keV.  Weighted mean colors for the MSPs in 47~Tuc
with and without 47~Tuc-J (labeled separately) are shown as the solid
square and triangle, respectively. }
\vspace*{0.14in}
\end{figurehere}

Several constraints on MSP x-ray emission may be inferred from
Figure~\ref{fig:cc}. 
Given the average MSP luminosity (from Table 1) of \Lx \about 2 \X
10$^{30}$ \lcgs, then for a BB with the suggested kT \about 0.22\,keV 
or T \about 2.6 \X 10$^6$K, the bolometric luminosity is 
L$_{x-bol}$ \about 3 \X 10$^{30}$ \lcgs and 
the total emitting radius is only r$_x$ \about0.2\,km. 
NS H-atmosphere models, as needed 
for thermal spectra of the quiescent LMXBs in 47~Tuc (cf. Heinke 
et al. 2002 and references therein), would decrease the 
temperature by a factor of \about2 and thus increase the 
emission radius to \about0.8\, km. 
For a dipole field, both thermal (isotropic) polar caps are likely visible
and the effective radius of each would be \about0.14\, km, 
or \about0.56km for H-atmosphere models. Such a small 
emission area strongly suggests emission from a radiating area 
much smaller than the entire polar cap, which has expected 
radius of r$_{pc}$ = ($\Omega$R$_*$/c)$^{1/2}$R$_*$ for a MSP with angular 
rotation frequency $\Omega$ and NS radius R$_*$ (or r$_{pc}$ = 2.6km 
for a 3\,msec period MSP and R$_*$ = 10km). 
Partially filled polar cap thermal emission  
(i.e. r$_x$ $<$ r$_{pc}$) 
was also found by Zavlin et al. (2002) for the bright MSP J0437$-$4715 
(for which a two-component model could be fit to the much 
better statistics data than possible for the 47 Tuc MSPs) and 
is predicted by recent polar cap heating models of 
Harding \& Muslimov (2002; hereafter HM02), 
although we find important differences (cf. \S 5). 
The small emission area implies x-ray pulsations 
(which cannot be measured given the 3.2\,sec ACIS-I integration
time) with a sinusoidal pulse shape appropriate to the fractional
visibility of the isotropically radiating thermal polar cap(s).  
In contrast, the narrower pulse duty cycles of \about10\% for some field
MSPs (and one in the globular cluster M28; cf. BT99) are probably due
to non-thermal beamed emission.

If the x-ray emission were a (pure) TB spectrum with kT \about1\,keV,
the required emission measure EM \about 4 \X 10$^{53}$cm$^{-3}$ would
suggest plasma densities n \about 3 \X 10$^{26}$R$^{-3/2}$\,cm$^{-3}$
for an emission region 
(assumed spherical)
of radius R. Such a region must not dominate the
observed dispersion measure of the MSP, which is typically DM = 24.4
cm$^{-3}$pc for the 47~Tuc MSPs with a variation $\delta$DM \about
0.1\,cm$^{-3}$\,pc that is likely due to 
the ionized
ISM in 47~Tuc (Freire et al.~2001b). Thus,  the MSP local
contribution to DM is probably n $\cdot$ R \lsim 0.03 cm$^{-3}$pc
\about1 \X 10$^{17}$ \cmsq.  Combined with the constraint above for EM,
for a constant density emission region this implies an unrealistically
large lower bound for the emission region: R \gsim 3pc. (For a TB
source with radius R \about 10$^{10}$ cm, or comparable to the size of
the typical MSP binary separation, n \about 3 \X 10$^{11}$ \cm-3 is
required which would strongly violate the DM constraint.)  Thus 
a pure TB spectrum is ruled out.  Similarly, with the exception of
47~Tuc-J, PL models are unlikely since their photon index must be implausibly 
large (\gsim3) and are still inconsistent with the colors
for most of the sample.  The limited statistics can not, of course,
rule out combined BB + PL models, but certainly it appears that the MSP
sample in 47~Tuc is dominated by thermal emission from polar caps of the
NSs.
 
\section{X-RAY VS. RADIO PROPERTIES OF MSPs IN 47~TUC}
The 47~Tuc MSPs are the first significant sample of MSPs at known
distance with measured x-ray and radio properties.  Several
long-standing pulsar questions can be addressed.

\subsection{X-ray vs. Pulsar Spin Properties}
Foremost is the correlation of x-ray luminosity and pulsar spindown
luminosity $\dot E$, which is found for field MSPs (with much more
uncertain distances) to scale as \Lx \about 10$^{-3}$ \Edot (BT99).
Here the globular cluster is both a hindrance and an interesting
laboratory: the MSPs are accelerated in the cluster potential, thus
biasing the observed values of \Pdot and hence \Edot ($\propto \dot
P/P^3$).  FCL01 have modeled the variations of \Pdot among the 15 MSPs
with precise locations (not including 47~Tuc-S) to derive constraints on
both the cluster potential and the intrinsic $\dot P_i$ values for the
MSPs. Tighter constraints on $\dot P_i$ can be obtained by assuming the
cluster gas (detected from variations in DM) is uniform so that the
observed DM value and projected radius give the 3D position in the
cluster for each MSP (see Freire et al.~2001b).  Assuming a King model for the 
cluster,
and using the central velocity dispersion \vdisp = 11.6\,km\,s$^{-1}$
(Meylan \& Mayor 1986), then gives the cluster acceleration term (cf.
Figure ~2 of Freire et al.~2001b) which is subtracted from the observed $\dot 
P/P$ to
infer the intrinsic $\dot P_i$.  Using a standard NS moment of inertia
$I=10^{45}$\,g\,cm$^2$, we then derive \Edot  = $4\pi^2 I \dot P_i/P^3$
for each MSP. Results are given in Table~\ref{tab:results} and plotted
vs.  \Lx in Figure~\ref{fig:lxedot}. Uncertainties in the derived 
\Edot values are typically 0.2--0.5 in the log (though several are 
larger) but are not shown for clarity; uncertainties in log(L$_x$) 
are typically 0.2 (see Table 1). For comparison with all MSPs previously 
detected in x-rays, we also plot values for 
10 field MSPs and one each in the globular clusters M28 and NGC~6397. 
The data for the field MSPs and M28 (from 
which \Edot values are derived) are summarized in 
the new compilation of data and distances in Table 2 
whereas the NGC~6397 x-ray results are from GHE01b and are discussed 
in detail in \S 4 below. 
We have repeated this analysis for variations in
both cluster distance (5.0\,kpc; +1.5$\sigma$) and  central velocity dispersion
(13.0\,km\,s$^{-1}$; +1$\sigma$) to sample the range of cluster acceleration
models, and thus intrinsic $\dot P_i$ values. Our goal is to derive the
approximate index $\beta$ in the relation \Lx $\propto$
\Edot$^{\beta}$.

\begin{figurehere}
\epsfig{file=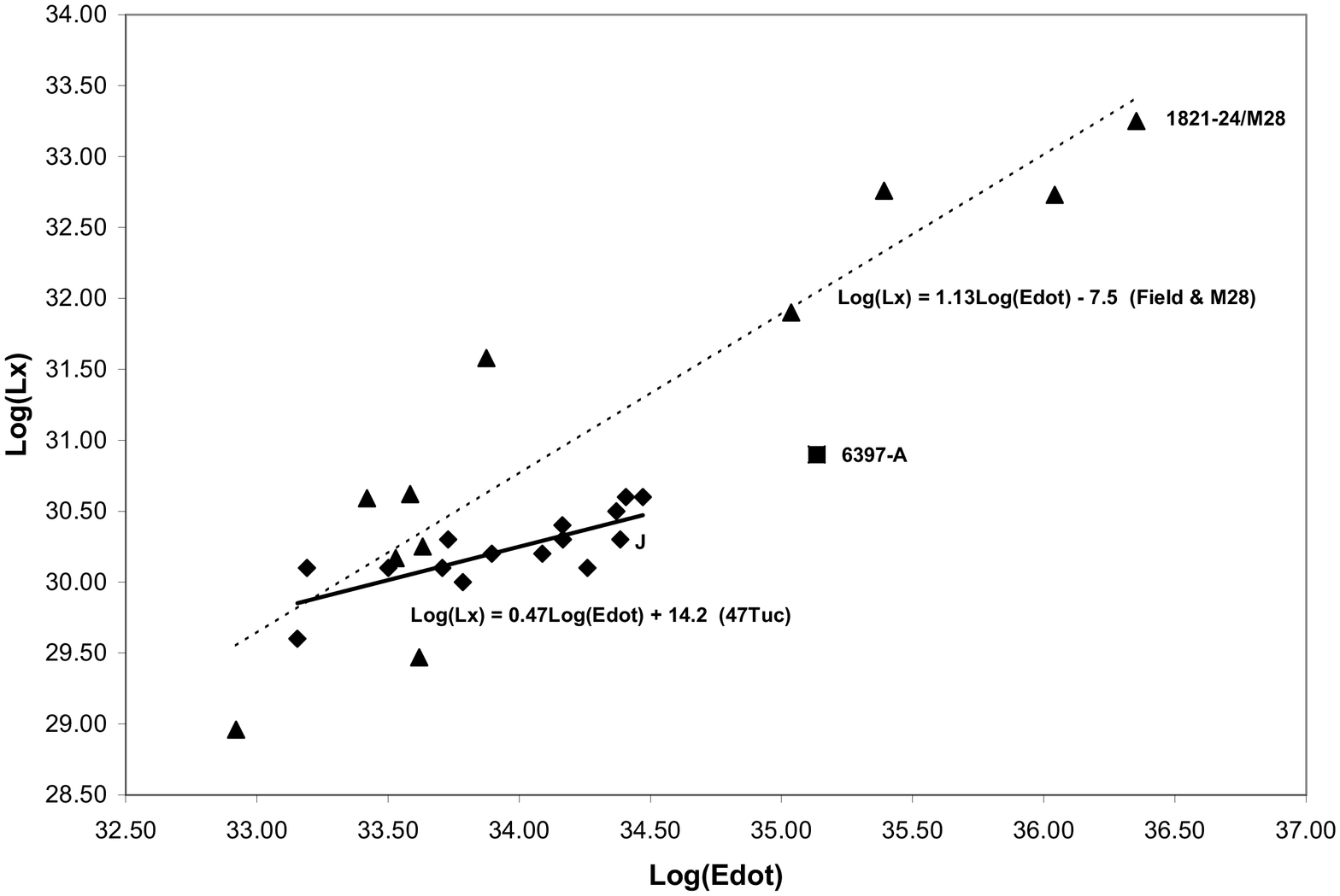,width=2.2in,angle=-90.}
\caption{\label{fig:lxedot} \Lx vs. \Edot for MSPs in 47~Tuc
(diamonds) and
field (triangles).  MSP in NGC~6397 (square) 
is labeled as is -J in 47~Tuc.  Values
for MSPs in field and M28 are from Table~\ref{tab:mspdata}, with \Lx =
$4\pi d^2 F_x$ and \Edot = $4 \pi^2 I \dot P_i/P^3$. }
\vspace*{0.14in}
\end{figurehere}

Whereas the field MSPs (and pulsars generally) are roughly consistent
with the linear relation \Lx \about10$^{-3}$ \Edot (BT99), the MSPs in
47~Tuc appear to have a much weaker dependence:  log\Lx =
($0.47\pm0.10$)log\Edot + $14.2\pm3.4$ for the nominal cluster model
with \vdisp = 11.6\,km\,s$^{-1}$, where the errors ($\pm$1$\sigma$) 
are due to just the
scatter in the points. Including the errors on each point, but with the
approximation that unequal errors (on $\dot E$) are simply averaged
(which biases the slope to steeper values, since the unequal errors are
much larger for smaller values of $\dot E$), increases the slope to
$0.59\pm0.28$ and offset to $10.0\pm10.6$. The best estimate of the
slope is thus $\beta$ \about$0.5\pm0.2$.  Allowing for the error in the
cluster model by using the +1$\sigma$ value for the central velocity
dispersion, \vdisp = 13.0\,km\,s$^{-1}$, and again including the errors
on each point gives slope $0.48\pm0.21$ and intercept $13.8\pm7.5$, and
so is consistent. Similar variations in $\beta$ and the 
intercept are found for the increased 
(by about 1.5$\sigma$) cluster distance, 5kpc. 
Thus, the MSPs in 47~Tuc have \Lx(0.5--2.5\,keV) 
values within a factor of \about4 (apart from 47~Tuc-C) despite a range of
\about25 in $\dot E$. For comparison, the field (and M28) MSPs in 
Figure~\ref{fig:lxedot} have $\beta$ = 1.13$\pm$0.15 and offset 
$-7.48\pm5.03$ and so have $\beta$ consistent with both the ROSAT 
band (0.1--2.4\,keV) results of BT99 and the 2--10\,keV results derived 
by Possenti et al. (2002), who find $\beta$ = 1.37$\pm$0.10. 

It is tempting to speculate these different $\beta$ values may be due to
the different formation histories and possibly different physical
parameters of MSPs in the field and 47~Tuc. In Figure~\ref{fig:lxage}a we plot
\Lx vs. spindown ages, $\tau$ =  $P/2\dot P_i$, for field and 47~Tuc MSPs. Error 
bars on the age parameter are not shown, for clarity, but are typically 
$\sim ^{+0.3}_{-0.1}$ in the log and primarily due to the cluster 
acceleration model. Despite the uncertainties, the correlation
is striking: the field MSPs show a declining \Lx with ``age,'' and the
47~Tuc MSPs appear to fall on this trend but 
with a flatter \Lx vs. age slope. 
(Although both age and $\dot E$ are derived from combinations of $P$ and 
$\dot P_i$, they are algebraically different so that the same flatter 
correlations with \Lx are not inevitable.)
In \S 5 we compare the \Lx/\Edot--$\tau$ relation (Fig.~4b) 
with the models of HM02, finding both qualitative 
similarities (noted by HM02) but also important differences.

\begin{figurehere}

\epsscale{.40}
\vspace*{0.1in}
\hspace*{-0.1in}
\epsfig{file=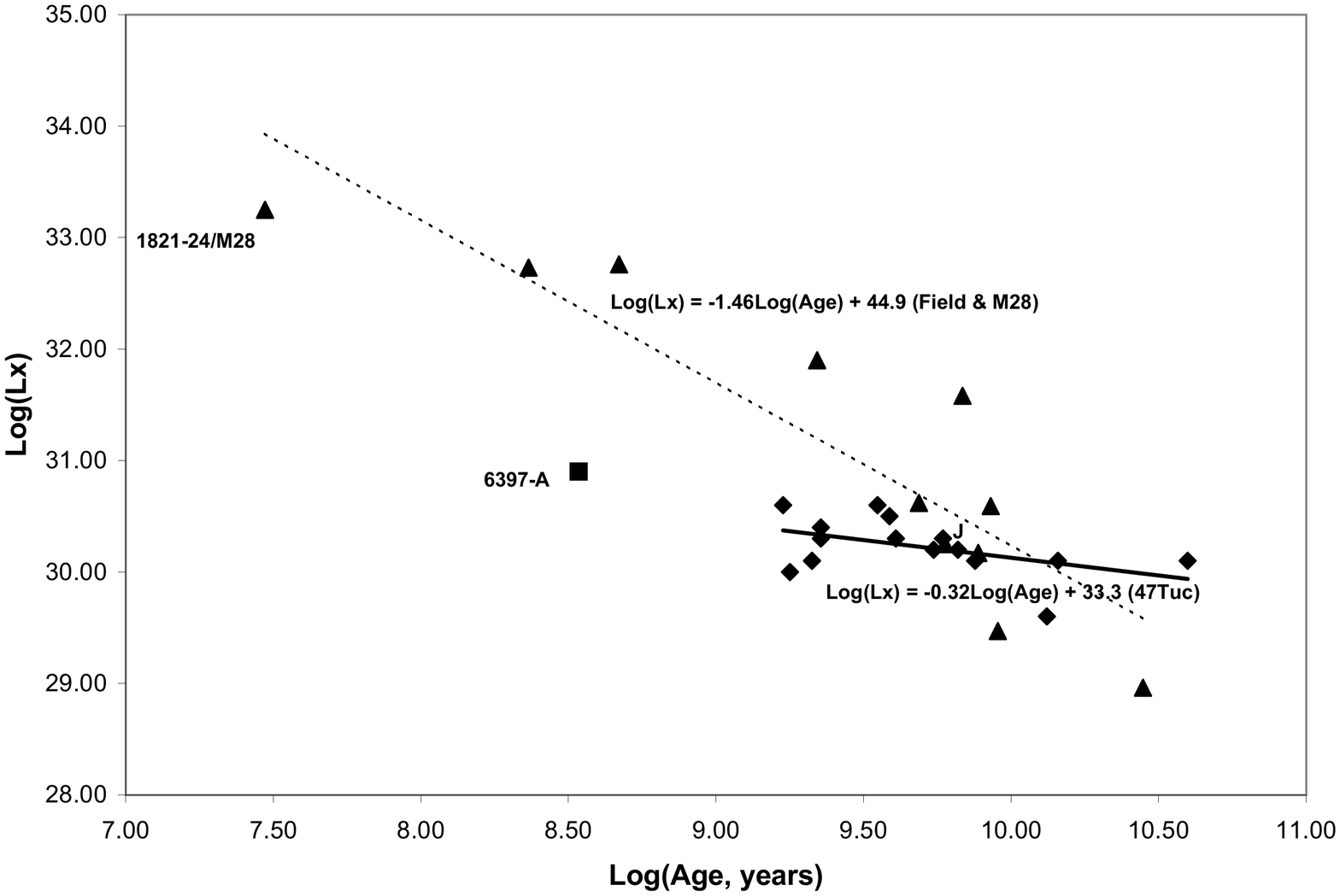,width=2.3in,angle=-90.}

\vspace*{0.1in}
\epsscale{.40}
\hspace*{-0.1in}
\epsfig{file=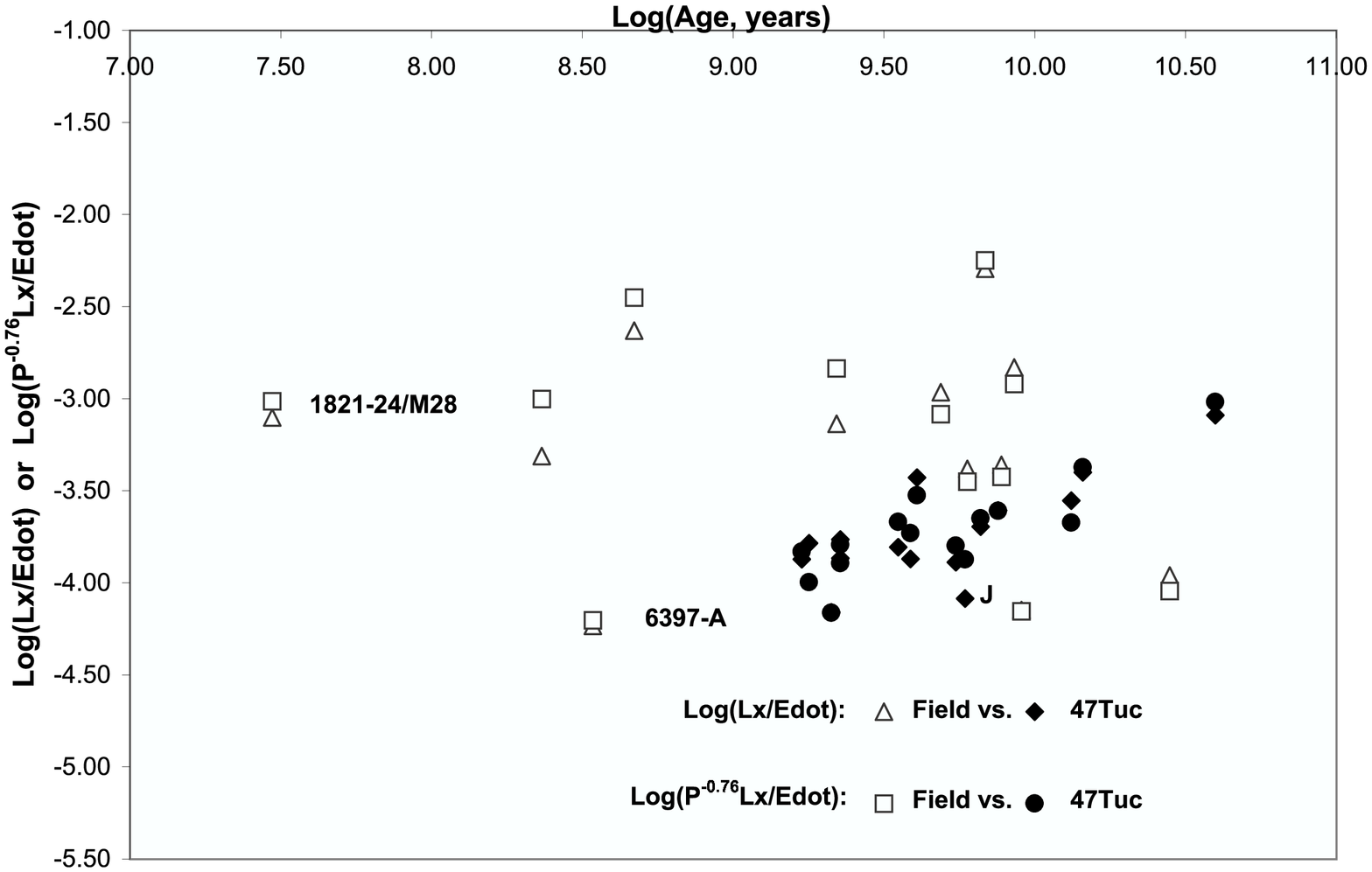,width=2.2in,angle=-90.}


\caption{\label{fig:lxage} Top (a): \Lx vs. spindown age ($\tau = P/2\dot P_i$)
for MSPs in 47~Tuc (diamonds) and field (triangles).  MSP in
NGC~6397 (square) is labeled as is -J in 47~Tuc.  
Values for MSPs in field and M28 are from
Table~\ref{tab:mspdata}. Bottom (b): \Lx/\Edot vs. spindown age, and
approximate correction for pulse period $P$ (from HM02; see text for 
empirical dependences on $P$ and $\tau$) for 47~Tuc vs. 
field (and NGC 6397) MSPs. }
\vspace*{0.14in}
\end{figurehere}

We note that although spindown ages 
correspond only approximately to actual ages, the derived 
values are consistent with the formation of most 47~Tuc 
MSPs in a ``burst" early in the cluster
history. However this by itself would not provide an explanation for the
different correlations observed for \Lx in 47~Tuc and field MSPs. 
As a measure of the systematic uncertainties in age, the 0.6\,Gyr value 
derived from the cooling age for the He-WD companion of 47~Tuc-U 
(Edmonds et al.~2001) may be compared to the \about0.9\,Gyr or 2\,Gyr 
spindown  age of the MSP (two solutions for $\dot P_i$ are possible 
for 47~Tuc-U, as is also the case for -G and -M 
from the total sample of 15 for which the cluster acceleration model 
yields values for $\dot P_i$ and thus \Edot and $\tau$). 

Another possible physical difference between the 47~Tuc and field MSPs 
might be magnetic field strength at the NS surface 
and thus also the light cylinder (at which the 
corotation speed equals $c$), since this likely affects the relative 
importance of non-thermal (magnetospheric) emission. 
For an assumed dipole field, this is given by 
$B_{\rm lc} = 9.35 \times 10^5 \dot P_{i}^{1/2} P_{\rm msec}^{-5/2}$ G, 
with $\dot P_{i}$ in units of 10$^{-20}$ s s$^{-1}$. 
In Figure~\ref{fig:lxblc}a we plot for the same pulsars \Lx vs. $B_{\rm lc}$. 
As before, considered as a homogeneous group, the field pulsars (and M28) 
lie on a 
significantly steeper logarithmic slope (1.70$\pm0.25$) than the 
47~Tuc MSPs (0.66$\pm0.13$). 
In contrast the field at the NS surface,  
$B_{\rm surf} =  3.2 \times 10^{19}$ ($P\dot{P}$)$^{1/2}$\,G, is 
nearly independent of \Lx. The correlation of \Lx with $B_{\rm surf}$ 
(Fig.~\ref{fig:lxblc}b) is less defined, 
with twice the scatter for both field and 47~Tuc MSPs, and   
with logarithmic slopes 
differing even more: 0.05$\pm0.27$ for the 47~Tuc MSPs 
vs. 2.80$\pm0.99$ for the field MSPs. 
Whereas at the light cylinder the dipole component of the 
field, $B_{\rm lc}$, would dominate, the surface field would 
be sensitive to multipole components which could give the 
larger scatter in Figure ~\ref{fig:lxblc}b. We return in \S 5 to 
additional arguments for a modified surface field topology. 

\begin{figurehere}

\epsscale{.40}
\epsfig{file=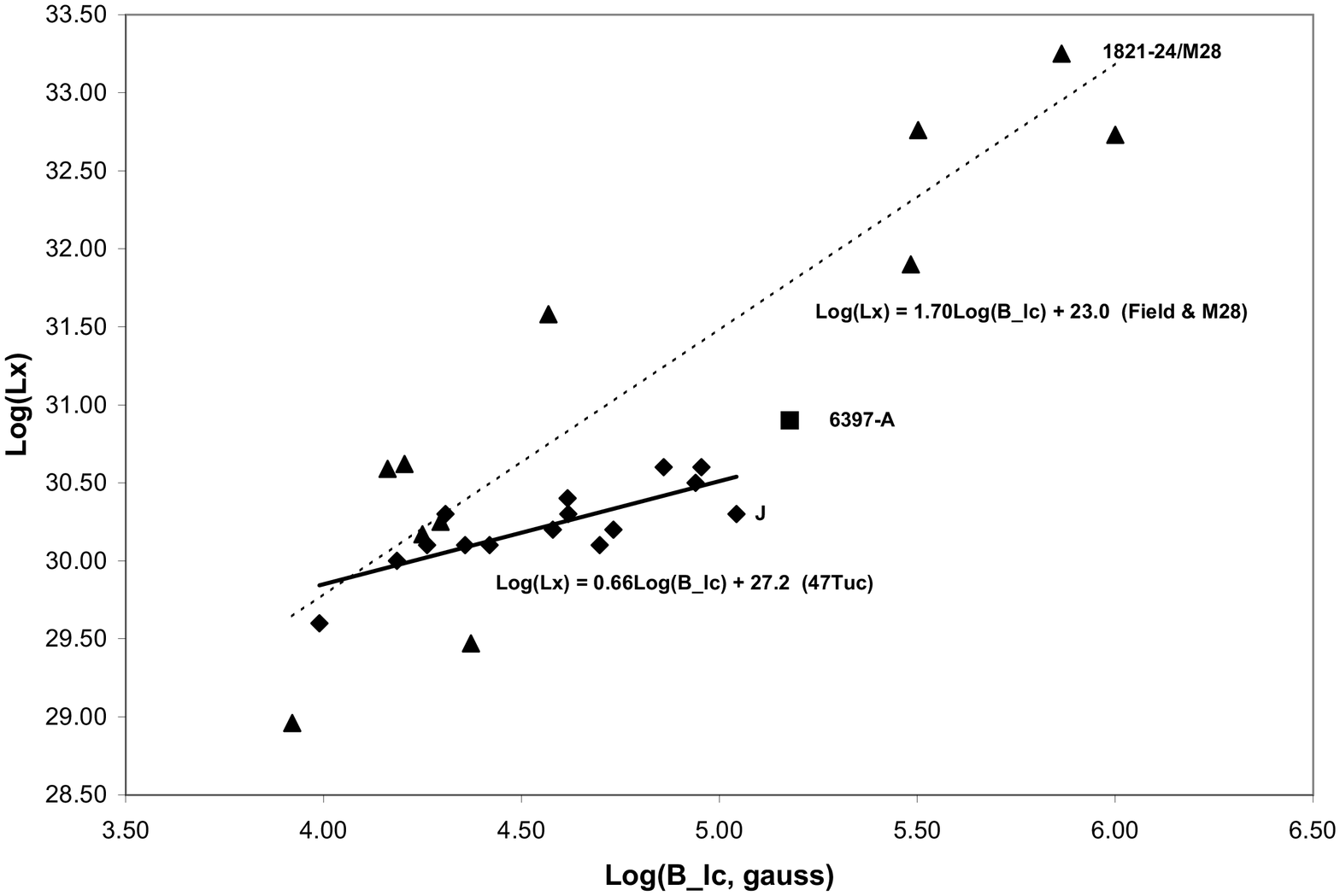,width=2.2in,angle=-90.}

\epsscale{.40}

\epsfig{file=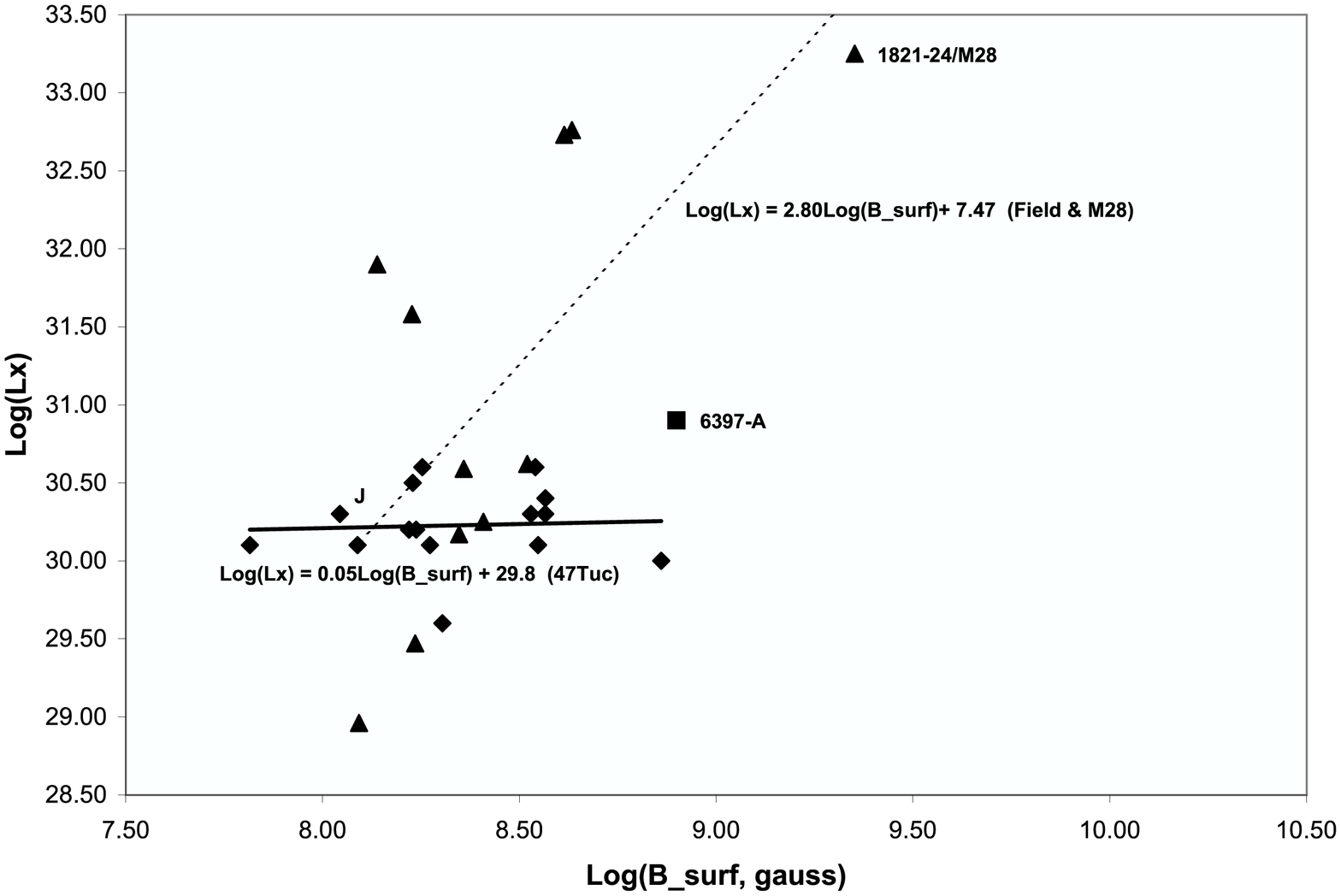,width=2.2in,angle=-90.}

\caption{\label{fig:lxblc} Top (a): \Lx vs. $B_{\rm lc}$, inferred magnetic
field strength at the light cylinder ($r_{\rm lc}=c P/2\pi$) for MSPs
in 47~Tuc (diamonds) and field (triangles).  MSP in NGC~6397 (square) is 
labeled as is -J in 47~Tuc.  Values for MSPs in field and M28 are from
Table~\ref{tab:mspdata}, with $B_{\rm lc}$ derived assuming standard
$1/r^3$ dependence for a dipole field (see text). Bottom (b):  \Lx vs. 
surface magnetic field (see text).}
\vspace*{0.14in}
\end{figurehere}

We note that 3 out of the 4 MSPs in the field 
with $B_{\rm lc} \ga 10^{5.5}$\,G display x-ray emission that is seemingly
magnetospheric, with the nature of emission from the 4th (the eclipsing
MSP B1957+20) indeterminate (BT99; Takahashi et al.~2001).  Conversely,
field pulsars with $B_{\rm lc} < 10^{5}$\,G have x-ray emission that is
typically either thermal or of indeterminate character (BT99).
Considering the small numbers of such pulsars studied, and that most of
them have highly uncertain distances (see Table~\ref{tab:mspdata}), it
is possible that field pulsars with $B_{\rm lc} \la 10^{5}$\,G may
show an L$_x$--$B_{\rm lc}$ trend that is fairly flat and roughly
consistent with the better determined relation for the 47~Tuc MSPs. 
However for this interpretation to hold, a few field MSPs with 
well determined distances (e.g. J0437$-$4715 and J1744$-$1134; cf. 
Table~\ref{tab:mspdata}) must be accounted for and the even larger 
deviations of J0751+1807 and J1024$-$0719 from the 47~Tuc correlation line 
would require a factor \gsim3 adjustments in these MSP distances. 
The fact that 5 of the 
7 field MSPs with log($\dot E$) \lsim 34 have 
\Lx values above the trend-line for the 47~Tuc MSPs may suggest these 
field MSPs still have non-thermal emission components. 
We return (in \S\S 5 and 6) 
to further discussion of the possibility 
that the magnetic field configuration and evolution for old cluster 
NSs may differ from those in the field.

\subsection{Radial Distributions} 
FCL01 pointed out the puzzling fact that the radial distribution of the
(then) 15 MSPs with timing positions seemed to be truncated within
\about3.5$r_c$, where \rcore = $24''$ is the core radius of the
visible stars in the cluster as determined most recently (and
completely) by Howell, Guhathakurta, \& Gilliland (2000). The addition
of 47~Tuc-S, as the 16th MSP with a precise location (Freire 2001), does
not change this.  It is therefore of particular interest to compare the
radial distribution of the 16 MSPs with the Chandra source distribution
over a region large enough to measure the radial profile of the sources
most likely to be MSPs. Our $2' \times 2\farcm5$ central field study
(GHE01a) suggested most of the MSPs are soft and a possible total MSP
population of \about50 in this region alone. Here we extend the study
of the Chandra sources out to a radius of 4\arcmin (\about10\,$r_c$),
to search for possible cutoffs or truncations in any of the Chandra
source distributions as might be expected if the MSP distribution were
cut off. We compare the MSP radial profiles with those of Chandra sources in
several apparent spectral (hardness) classes to further test the
possible identification of faint soft sources (``red background
sources'') with the large MSP population suspected (Camilo et al.~2000;
GHE01a).  This analysis will include, then, the vast majority of sources
associated with 47~Tuc. Complete details (spectra, timing) of this
source population will be included in the still-larger study (full ACIS
16\arcmin field) of 47~Tuc to be reported by Heinke et al. (in
preparation).

\begin{figure*}
\vspace*{-1.0in}
\epsfig{file=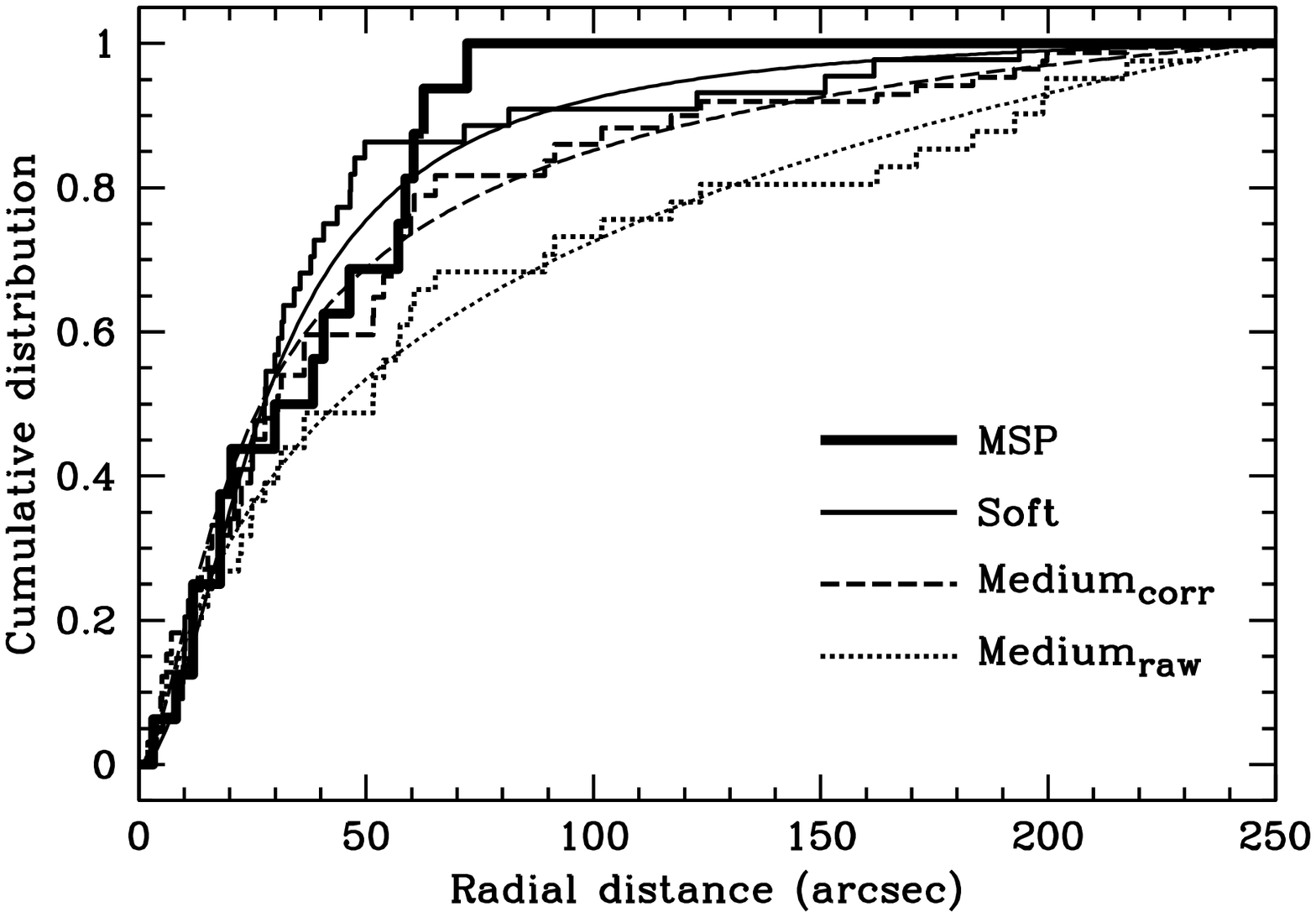,width=3.5in}
\vspace*{-1.0in}
\epsfig{file=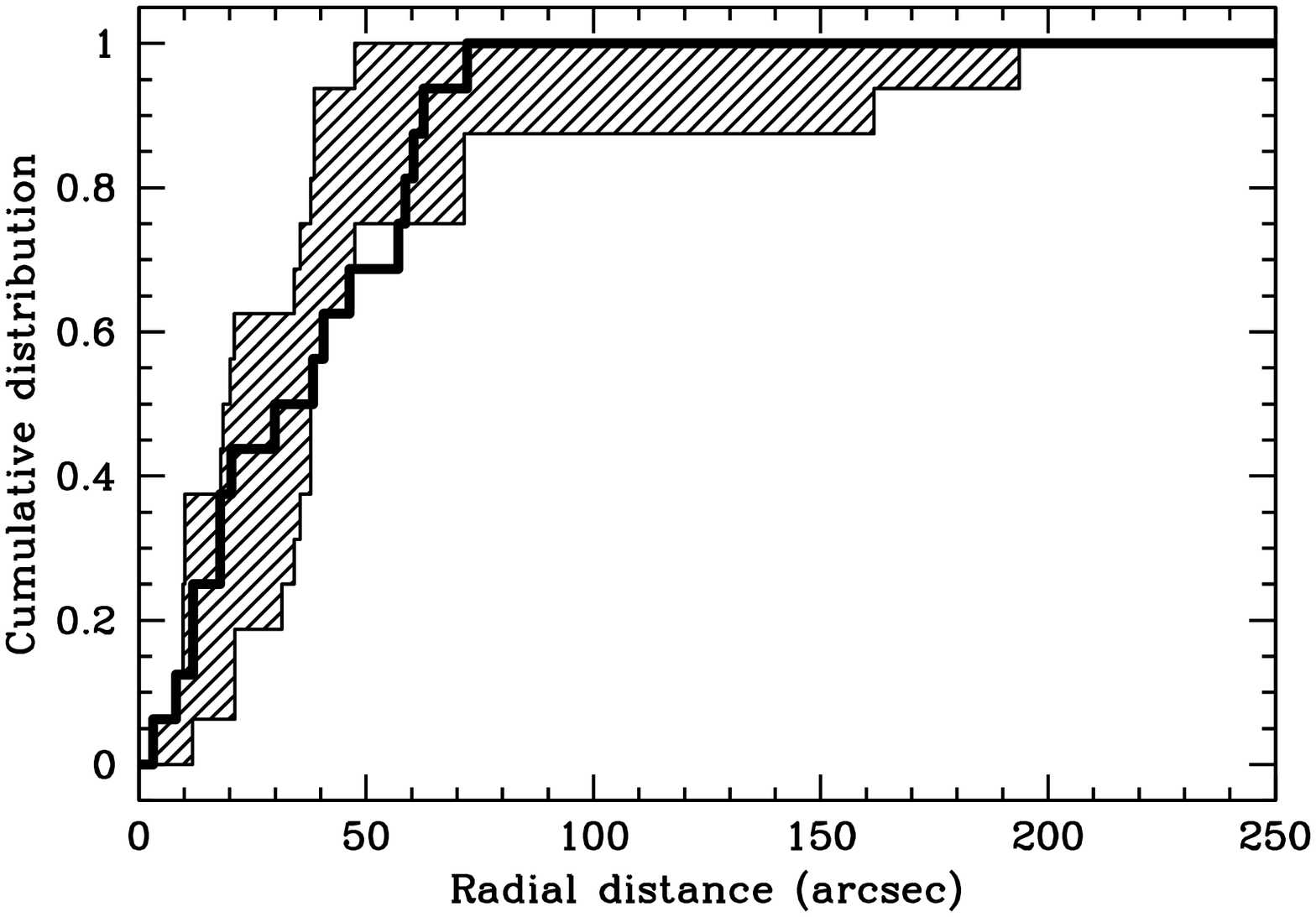,width=3.5in}

\vspace*{0.9in}

\caption{\label{fig:radial} Top (a): radial source distributions for
47~Tuc MSPs vs.  Chandra Soft (not including MSPs) and Medium (raw and
with statistical subtraction of expected AGNs) sources detected with
\gsim5cts together with best-fit King model predictions for each
group.  Bottom (b):  envelope (5\% and 95\%) of radial distributions for
1000 samplings of 16 soft Chandra sources (hatched region) vs. the MSP
sample (solid histogram). }
\end{figure*}

\WAVDETECT analysis of the the 4\arcmin (radius) field yields a total
of 187 sources above a limit of medcts(0.5--4.5\,keV) = 3 (vs.  103 in
GHE01a within the central $2' \times 2\farcm5$ box), or 126 sources above
a 5ct limit 
(vs. 92 in the $2' \times 2\farcm5$ box),
which we impose to get a $\geq3\sigma$ significance. 
For softness ratio (Xcolor =
2.5log[Soft(0.5--1.5\,keV)/Hard(1.5--6.0\,keV)]; cf.  GHE01a)
determination, we impose the \WAVDETECT
count limit of 3cts as an upper limit in either Soft or Hard bands if
there is no \WAVDETECT detection in that band. This alters the Xcolor values
somewhat (for weak sources) from those given in GHE01a.  
In Figure~\ref{fig:radial}a we provide the radial profiles of the
sources detected with \gsim5cts in at least one band, which are divided
into Soft (Xcolor \gsim 1.0; 44 sources) and Medium ($-$1.0 \lsim
Xcolor \lsim1.0; 41 sources) groups. For the Medium group we also
required that the source be detected by \WAVDETECT (\gsim 3 cts) in
both the soft (0.5--1.5\,keV) and hard (1.5--6.0\,keV) bands.
Not plotted are the Hard (Xcolor
\lsim$-$1.0; 9 sources) sources and remaining sources (32) not included 
in the above 3  groups, since
the Hard sources likely have a significant background AGN contribution
(though at least 2 of these 9 sources are cluster CVs; GHE01a), and the
remaining sources are indeterminate.  The Medium source group, and thus its
radial distribution, is also likely to be contaminated in this larger
field by background AGNs. Scaling from the deep survey counts and AGN
colors measured by Giacconi et al.~(2000), we expect \about8 Medium and
\about3 Hard sources to be background AGN.  Therefore, the Medium
source distribution has also (Fig.~\ref{fig:radial}a) been corrected by
the following Monte-Carlo procedure.  A total of 1000 bootstrap
resamplings of the distribution were generated.  For each of these
samples, an expected number of background sources was selected from a
Poisson distribution with mean 8.  This number of radial source
positions was then selected from a uniform spatial distribution
(i.e.\ with cumulative distribution $\propto r^2$), and the closest
actual sources to these positions were removed from the sample.  The
average of the 1000 background-corrected sample distributions was then
adopted as the best estimate for the corrected cumulative
distribution.

In Table~\ref{tab:ks} we give the probabilities from a KS test that
each source group has a radial distribution consistent with that of any
other group (including the 16-member MSP sample).  For the Medium
group, we show results for both the raw and corrected distributions.
All of the cross-sample comparisons indicate consistency
(i.e.\ probability of the samples being consistent exceeds 5\%) with
the exception of the Medium-raw -- Soft case.  We ascribe this
inconsistency to the contamination of the Medium-raw sample by
background sources and note that the Medium-corrected sample is
consistent with the Soft sample.

In Figure~\ref{fig:radial}b we show that 1000 repeated samplings of 16
sources from the Soft population gives a range of radial distributions
that includes the observed MSP distribution.  We note that both the CVs
and BY Draconis (BY Dra) main sequence binaries have radial
distributions that are also consistent with that of the Soft sources.
However the radial incompleteness of our present HST sample and thus CV
and BY Dra identifications makes this result preliminary.

For each of the observed or corrected (Medium) radial distributions in
Figure~\ref{fig:radial}a, we have fitted what we shall refer to as a
``generalized King model'' profile (cf.\ Lugger, Cohn, \& Grindlay
1995) to the projected surface density,
\begin{equation}
S(r) = S_0 \, \left[1 + \left({r \over r_0 }\right)^2 \right]^{\alpha/2}.
\end{equation}
Here $\alpha$ is the large-$r$ power-law index and the core radius
$r_c$ is related to the radial-scale parameter $r_0$ by
\begin{equation}
r_c = \left(2^{-2/\alpha} - 1 \right)^{1/2}\,r_0.
\end{equation}
The corresponding space density profile is given by
\begin{equation}
n(r) = {\Gamma(\frac{-\alpha}{2}) \over \sqrt{\pi}\,
\Gamma(\frac{-\alpha-1}{2})} \, {S_0 \over r_0} \,
\left[1+\left({r \over r_0}\right)^2 \right]^{\frac{\alpha-1}{2}}.
\end{equation}

Following the analyses of Jernigan \& Clark (1979), and Lightman,
Hertz, \& Grindlay (1980), we assume that the projected distribution of
the dominant 
visible stellar population, with mass $M_*$ ($\sim0.7\,M_\odot$ in
the cluster core due to mass segregation), 
is described by an ``analytic King model,''
i.e.\ the form given by eq.~(1) with $\alpha_* = -2$.  In thermal
equilibrium, the space density of an x-ray source population with
stellar mass $M_x = q M_*$ is given by $n_x(r) \propto n_*(r)^q$.
Thus, the power-law index of the projected x-ray source density profile
is predicted to be
\begin{equation}
\alpha_x = -(3q-1)
\end{equation}
and the corresponding core radius is predicted to be
\begin{mathletters}
\begin{eqnarray}
r_{cx}  & = & \left( 2^{2/(3q-1)} -1 \right)^{1/2}\, r_{c*}  \\
         & \approx & 0.68 \left( q - \frac{1}{3} \right)^{-1/2}\, r_{c*}.
\end{eqnarray}
\end{mathletters}%
The latter approximation holds in the limit $q\gg1$ and is accurate to
7\% for $q\ge2$.

In Table~\ref{tab:king} we list the values derived for $r_c$ and
$\alpha$ for each source group.  The Soft sample is of particular
interest, since it is the largest one ($N=44$) and its parameters are
reasonably well determined.  As we have shown in Table~\ref{tab:ks},
both the Medium-corrected and the MSP samples are consistent with the
Soft sample.  In principle, the mean x-ray source mass $M_x$ may be
computed from eqs.~(4) and (5).  Eq.~(4) gives $q = 1.5$ for the
best-fit value of $\alpha_x = -3.6\pm0.8$, with a 1-$\sigma$
uncertainty range of $q = 1.3$--1.8.  Eq.~(5a) gives $q = 1.7$ for the
best-fit value of $r_{cx} = 15\farcs2\pm3\farcs5$ together with $r_{c*}
= 24\farcs0$ (Howell et al.\ 2000), with a 1-$\sigma$ uncertainty range
of $q = 1.3$--2.5.  Thus, both measures of $q$ put the best-fit value
at around 1.6 and allow values as large as $q\approx2$.  Given our
estimate that the dominant mass in the core of 47~Tuc is $M_* \sim
0.7\,M_\odot$, this implies that $M_x$ lies in the range $\sim
1.1$--1.4$\,M_\odot$.  For comparison, Grindlay et al.\ (1984) obtained
a most probable value of $q=2.6$ for an ensemble of eight luminous
LMXBs in as many clusters.

There are a number of limitations to the ``classic'' analysis outlined
above, which may well affect the estimate of the x-ray source mass.
First, this analysis assumes that the potential is dominated by stars
of a single mass.  In fact, there is a large range of stellar mass with
substantial mass segregation.  Secondly, the analysis assumes that the
x-ray sources make a negligible contribution to the gravitational
potential.  However, given mass segregation, even a modest NS or heavy
white dwarf (WD) population can make a substantial contribution to the
potential in the innermost part of the cluster.  Finally, the analysis
assumes that the cluster is in a state of thermal equilibrium,
i.e.\ that the cluster is isothermal and that each species is in energy
equipartition.  This condition does not hold rigorously, even in a nearly
isothermal multi-mass King model. 
Also, the relatively low $M_x$ \about 1.1--1.4$\,M_\odot$ may suggest  
non-equilibrium (perhaps from binary scattering) if the NS masses are 
\gsim1.4 \Msun as possibly suggested by the long-lived thermal emission 
(cf. \S 5).

In light of these considerations, it is useful to compare our results 
for the Soft source distribution with the predictions from actual
dynamical models.  Meylan (1988) fitted multi-mass, anisotropic
King-Mitchie models to surface brightness and stellar velocities for
47~Tuc, in order to constrain cluster parameters including the mass
function.  The models include a NS component, which comprises between
0.1--0.5\% of the total cluster mass of $7\times10^5\,M_\odot$.  Radial
profiles for all components are given for one model with a NS mass
fraction of 0.2\% (equivalent to $\sim 1000$ NSs in the cluster).  The
power-law for the NS component has a slope of $-4.0$ in projection,
which is consistent with the value of $\alpha = -3.6\pm0.8$ found here
for the Soft source distribution. The predicted core radius for the NS
component is $16''$, which is also in good agreement with our
determination of $r_c = 15\farcs2\pm3\farcs5$ for the Soft sources.  By
comparing the observed Soft source distribution with a set of models
with a range of NS fraction, it should be possible to constrain this
parameter.  The models may be generated either by following Meylan's
(1988) King-Mitchie model approach, or by Fokker-Planck integrations
(e.g.\ Murphy et al.\ 1998).  We are undertaking a study based on the
latter method.

\subsection{Candidate MSP Sample in 47~Tuc and Estimated Total}
The KS probabilities (Table~\ref{tab:ks}) and King model fits
(Table~\ref{tab:king}) show that the radial distribution of the 44 Soft
sources is indistinguishable from that of the MSPs, and indeed 7 of the
9 sources identified with MSPs with well-measured Xcolor values are
soft. These soft MSPs include the doubles (-G, -I) and (-F, -S), for
which separate Xcolors could not be determined, so that in fact 9
MSPs are accounted for.  The remaining MSPs are either faint (-M, -T;
softcts \lsim5; cf. Table~\ref{tab:results}) or crowded (-L, -Q), but
it can nevertheless be seen from the x-ray colors plotted in
Figure~\ref{fig:cc} that all 47~Tuc MSPs, with the exception of 47~Tuc-J,
are probably Soft (although 47~Tuc-U, with Xcolor = 0.9, is formally in
the medium group, while 
colors for 47~Tuc-C could not be measured).

Motivated by these results we have searched the list of Soft sources
for extra MSP candidates. Of the 37 Soft sources not identified as
MSPs, two are likely qLMXBs (X5 and X7; GHE01a and Heinke et al.~2002),
two are CVs (V3 and AKO 9; GHE01a) and three are sources with counts more
than two times brighter than the brightest detected MSP. Only 17 of the
remaining 30 sources fall in the HST field of view (FOV) of Albrow et
al.~(2001).  Nine of these 17 sources are astrometrically matched to
optical variables 
(hereafter ``active binary" 
candidates) identified with HST
(Edmonds et al., in preparation), and an additional 2 are faint blue
stars and possible CVs, one of which is also a periodic variable.
Seven of the optical variables have independently been identified as
likely BY Dras, red stragglers, or contact/semi-detached binaries by
Albrow et al.~(2001), and 2 of them were reported by GHE01a as BY Dras
based on the additional (more restrictive) constraint of significant
x-ray variability.

Despite the lack of detection of any of the 7 soft MSPs as optical
variables, it is possible that several of the active binary candidates
contain MSPs. For example the optical companion to the one MSP known in
NGC~6397 was previously identified as a BY Dra candidate by Taylor et
al. (2001; hereafter TGE01). The corresponding Chandra source is
relatively hard (see \S~4) and would not have been included in our list
of soft sources in 47~Tuc, but we cannot be sure that all other MSPs
with main sequence 
(or red straggler, as likely for the NGC~6397 MSP) 
companions will be similarly hard. 
In fact, after the initial submission of this paper, 
the relatively hard Chandra source W29 (GHE01b) 
has been identified with the MSP 
47~Tuc-W as having a main sequence companion (Edmonds et al. 2002; cf. 
\S~4). 
Also, one or both
of the soft sources with a faint blue optical counterpart could be a He
WD companion to an MSP (like 47~Tuc-U; Edmonds et al.~2001), instead of
a CV.  The definitive MSP characteristic is, of course, a pulsed radio
source. 
However, 
a significantly larger sample of MSPs with radio 
timing positions, for comparison with the optical variables, will be
difficult to obtain because so much observing time has been devoted to
finding the 16 already known.  Therefore, to estimate the number of
MSPs implied by our soft sample we consider two extreme cases: (a) none
of the candidate optical counterparts (9 active binary and 2 CV
candidates) are MSPs, or (b) all of the candidate optical counterparts
are MSPs. For case (a) we find that only 6 of the soft sources falling
in the HST FOV (and with counts $<2\times$ the brightest detected MSP)
do not have a known ID.  Correcting for the detected fraction of MSPs
with timing positions that are soft sources (7/16) and the incomplete
FOV of HST (17/44 sources are outside the HST field) we estimate an
extra population of 19 ($=6[1+17/44]\times16/7$) MSPs with \Lx $\gtrsim
10^{30}$ \lcgs, in addition to the 16 already known. For case (b) a
total population of 69 ($=30\times16/7$) extra MSPs is implied, for a
total maximum MSP population of 85.

These estimates are based on scaling from the number of soft source
detections with \WAVDETECT to the total MSP population with timing
positions. Using a different technique we estimate the number of soft
sources missed by \WAVDETECT by examining the net counts {\em not\/}
attributable to \WAVDETECT sources within a radius of 24\arcsec
(=1\,\rcore for the background stars), where crowding will most limit
faint source and crowded-field detection, and the annulus 1--2\,\rcore
surrounding the cluster center. By removing counts from the original
image in 1\arcsec radius regions around each \WAVDETECT source (from
GHE01a) in both the Softcts and Mediumcts bands (cf.
Table~\ref{tab:results}), and then subtracting a fraction of the
remaining counts (in each band) to correct for the psf-spillover from
these bright(er) sources due to the energy-dependent Chandra telescope
psf, we find 180$\pm$56 Softcts, 20$\pm$58 Mediumcts and 15$\pm$36
Hardcts in the central 1\,\rcore radius region, and 60$\pm$24,
75$\pm$28 and 45$\pm$25, respectively, in the surrounding 1--2\,\rcore
annulus.  Thus, as suspected in GHE01a, there is an excess of Softcts in
the cluster core, but the uncertainties in this extraction are large
and 
no excess in the core of Mediumcts or Hardcts is  required.  
For  
the detected MSPs with Softcts as listed in Table~\ref{tab:results} (with
mean value 6cts), the excess total Softcts (240) 
could arise from an
additional \about40 MSPs, all with \Lx \gsim10$^{30}$ \lcgs.  Since
some of these soft sources 
could be BY Dras (coronal sources are
typically soft) or CVs, as already noted (GHE01a), we regard 40 as an 
upper limit and estimate \about20 (half) ``unresolved'' MSPs in the core.

Scaling by our known sample of 16 MSPs, we derive $40\times(16/7)=91$
MSPs as an upper limit on the total MSPs in the cluster.  This upper
limit agrees with the upper limit derived from consideration of 
\WAVDETECT soft
sources, and we adopt 90 as an approximate maximum number of MSPs 
in 47~Tuc with \Lx \gsim10$^{30}$ \lcgs.  
Both of these estimates assume that no beaming of the MSPs in
x-rays is occurring. 
(Note that thermal polar cap emission, while pulsed, is 
not beamed.) 
If there are more MSPs than this in 47~Tuc, they
must have 
relatively hard spectra (since the above totals are 
scaled from the soft source counts) which could make them 
like the MSP in NGC 6397 (see below) for which radio 
emission is occulted much of the time. Alternatively, additional MSPs are 
beamed away from us in the x-ray, or of significantly lower
x-ray luminosity on average.  The softer color of the excess counts
than our known MSP sample suggests that either the excess counts are
composed largely of other sources (such as faint BY Dra binaries), or
that there is a significant population of faint (in x-ray and radio)
MSPs which would have cooler (and thus softer) polar caps.

\begin{figurehere}
\epsfig{file=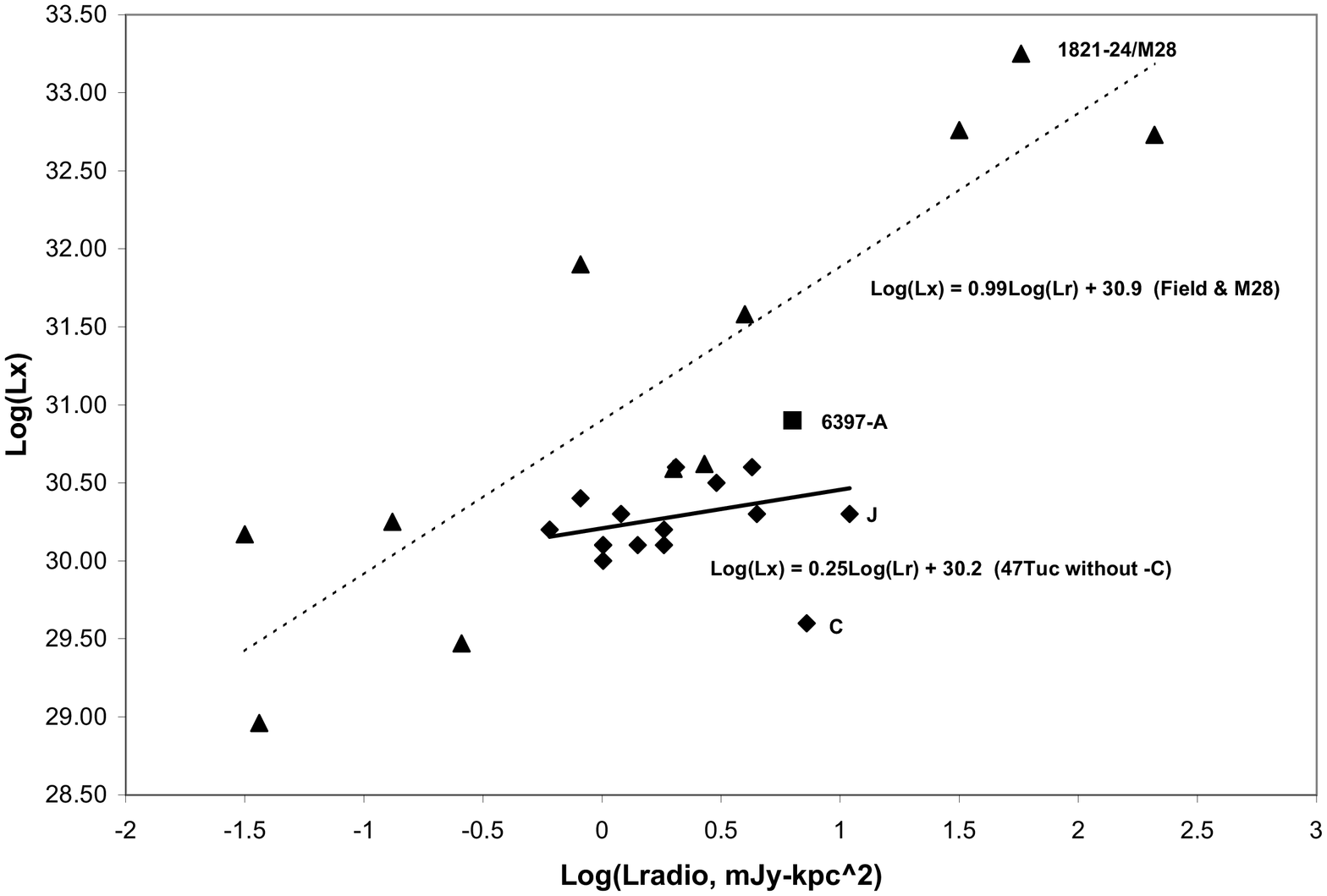,width=2.2in,angle=-90.}
\caption{\label{fig:lxlr} \Lx vs. L$_{\rm radio}$ for MSPs in 47~Tuc
(diamonds) and field (triangles).  MSP in NGC~6397 (square) is 
labeled as is -J in 47~Tuc.  Values for MSPs in field and M28 are from
Table~\ref{tab:mspdata}, where L$_{\rm radio} = S_{1400}d^2$.  47~Tuc
MSP $S_{1400}$ values are from Camilo et al.~(2000). Correlation line 
shown with 47~Tuc-C excluded (see text). }
\vspace*{0.14in}
\end{figurehere}

A significant number of additional (\gsim15--70) MSPs already detected
in the Chandra data but not yet detected in the deep Parkes radio
surveys is quite possible, given the relatively flat dependence of soft
x-ray luminosity, L$_x$, on effective radio luminosity, L$_{\rm radio}
= S_{1400}d^2$, where $S_{1400}$ is the flux density at 1400\,MHz and
$d$ the MSP distance. This relation is shown in Figure~\ref{fig:lxlr}
for 47~Tuc, NGC~6397, M28 and field MSPs.  The 47~Tuc MSPs have a
distinctly flatter L$_x$ dependence on L$_{\rm radio}$, 
with logarithmic slope 0.00$\pm$0.23 for all the 47~Tuc MSPs or  
0.25$\pm$0.15 with 47~Tuc-C excluded from the fit, 
as shown in Figure~\ref{fig:lxlr}. The logarithmic slope is 
0.99$\pm$0.17 for the field MSPs (and M28), which
is at least partly affected by sample selection bias.  Thus it appears 
that in 47~Tuc MSPs that are fainter at radio wavelengths are not as 
much fainter in x-rays and should be more readily detected in Chandra
data than in Parkes data. 
The ratio of 47~Tuc MSPs likely detected in x-rays 
(\about 35--90) vs. radio (20) 
can constrain the relative beaming fractions 
since both samples are detected above the same x-ray luminosity 
limit (\Lx = 10$^{30}$ \lcgs). 
If the x-rays are 
not beamed, the implied radio beaming fraction is \about 0.2--0.5.

We cannot (yet) isolate the candidate MSPs among the unresolved
component, but as a possible aid to future MSP identifications, we
report the positions and fluxes of the  19 Soft sources
detected which are not identified with possible BY Dra or CV
candidates.  In Table~\ref{tab:cands} we give the source positions,
detected counts, and approximate luminosities \Lx (again, assuming for
consistency a kT = 0.2\,keV BB spectrum in the 0.5--2.5\,keV band).
These are all possible MSP candidates (though some BY Dras and a few
CVs are likely included) and thus 
appropriate for use as trial positions in timing solutions for 
detected radio MSPs.

\section{COMPARISON WITH MSP IN NGC~6397}
We now compare the properties of the 47~Tuc MSPs with those derived for
the single MSP in NGC~6397. This MSP, PSR~J1740$-$5340 (D'Amico et
al.~2001a), hereafter 6397-A, was recently located by a pulse timing
solution (DPM01) and identified with an optical counterpart (Ferraro et
al.~2001)
previously identified with a main sequence binary or BY Dra system
(TGE01).  Chandra observations of NGC~6397 (GHE01b) detected 20 sources
within 1\arcmin of the cluster center, including the MSP. Because the
Chandra sources included 8 CVs firmly identified with HST optical
counterparts displaying \Halpha emission (GHE01b),  
the Chandra-HST boresight could be measured to \about
$0\farcs01$ in RA and Dec. Thus the astrometric identification of the
Chandra source U12 (GHE01b) with the apparent BY Dra binary WF4-1 found
by TGE01 is secure (the 
Chandra-HST positional offsets are given in
Table~\ref{tab:results}, and the Chandra-radio offsets are 
given in GHE01b). 
Since this binary has an optical photometric
period 1.34\,d (TGE01) which is consistent with that of 6397-A (D'Amico
et al.~2001a), and is included in the \about1\arcsec astrometric
uncertainty of the HST guide star vs. radio timing reference frames, it
was identified as the probable optical counterpart by Ferraro et al. (2001). 
In addition to the precise Chandra positional identification of the 
1.34\,d binary, 
we confirm the MSP identification by its x-ray temporal
variability which was mentioned by GHE01b as being consistent with
modulation at the binary period.  

In Figure~\ref{fig:6397} we show the x-ray light curve of 6397-A (source
U12) plotted against the binary phase as determined from the ephemeris
of DPM01. The count rate increases smoothly by a factor of \about2 at
phase \about0.4, or just before the MSP comes out of radio eclipse
(which is centered at phase 0.25; phase 0 is the time of ascending
node), removing any doubt that the MSP is the x-ray counterpart. The
fact that the flux is not zero during eclipse (though Chandra coverage
only begins at phase \about0.1) suggests that the x-ray source is
extended and larger than the companion size, which 
has radius \about1.3--1.8\Rsun and  nearly fills 
its Roche lobe (Burderi et al. 2002). 
Thus the x-ray emission is likely to be
from the relativistic wind from the MSP interacting with the mass loss
from the companion. 
Any thermal (BB) component from the MSP itself 
must be \lsim0.5 the
total flux, as constrained by the flux increase out of eclipse.  The
hardness ratios during (HR1=4.7$^{+5}_{-1.6}$, HR2=0.42$\pm{0.16}$) and
outside eclipse (HR1=3$^{+1.7}_{-0.9}$, HR2=0.57$\pm0.2$) remain the
same within the errors, implying 
the modulation is not due to 
increased $N_H$. More sensitivity and phase coverage is needed
to confirm and constrain the modulation. 

The x-ray counts detected in each of the Soft, Medium and Hard bands
are given in Table~\ref{tab:results}, together with the \Edot value
from DPM01, which assumes that, given its relatively large projected
offset of $0\farcm6$ (\about11\,$r_c$) from the cluster center, the
cluster acceleration may be ignored. MSP 6397-A is clearly very hard,
with hardness ratios HR1 = 3.6 and HR2 = 0.5. We do not plot it in
Figure~\ref{fig:cc} since the very different absorption column ($N_H$ =
1.0 \X 10$^{21}$ \cmsq, or \gsim4\X\ that for 47~Tuc) precludes direct
comparison with the models plotted for 47~Tuc, but it is consistent with
a PL with photon index \about1.5 or TB with kT \gsim5 keV.

However we can compare the derived \Edot and characteristic age, and
their relation to L$_x$, for this MSP with the 47~Tuc sample.  In
Figures~\ref{fig:lxedot} and \ref{fig:lxage} we plot 6397-A  
but do not include it with the linear fits which were derived
for the 47~Tuc MSPs alone.  It is clear that 6397-A is  consistent with
the 47~Tuc sample and correlations shown in both figures.  Including it
on the \Lx vs.  \Edot correlation would give a combined (47~Tuc and
NGC~6397) \Lx $\propto$ \Edot$^{\beta}$ relation with $\beta =
0.48\pm0.15$, even including the errors on all the points (despite the
bias towards steepening the slope in doing so, as mentioned before).
This is at first surprising, given the very different spectral shape
and thus emission region, as well as (much) smaller spindown age,
$\log(P/2\dot P) = 8.5$, for 6397-A.  However the eclipsing MSP in
47~Tuc with the hardest spectrum, 47~Tuc-J, already hints that the x-ray
spectral shape is not critical to the \Lx vs. \Edot correlation, since
-J does not depart from the correlation line. Thus there seems to be a 
relatively fixed \Lx(0.5--2.5keV) for both the 47~Tuc MSPs and 6397-A 
which emerges for a given spindown power \Edot regardless of the 
emission mechanism. It appears that the conversion of \Edot into thermal (BB) 
emission from the NS and emission probably tied to the  relativistic 
pulsar wind are both fixed by the total polar cap current which, from 
the models of HM02 is in turn proportional to \Edot$^{1/2}$ (see \S 5).  

\vspace*{0.1in}
\begin{figurehere}
\hspace*{0.1in}
\epsfig{file=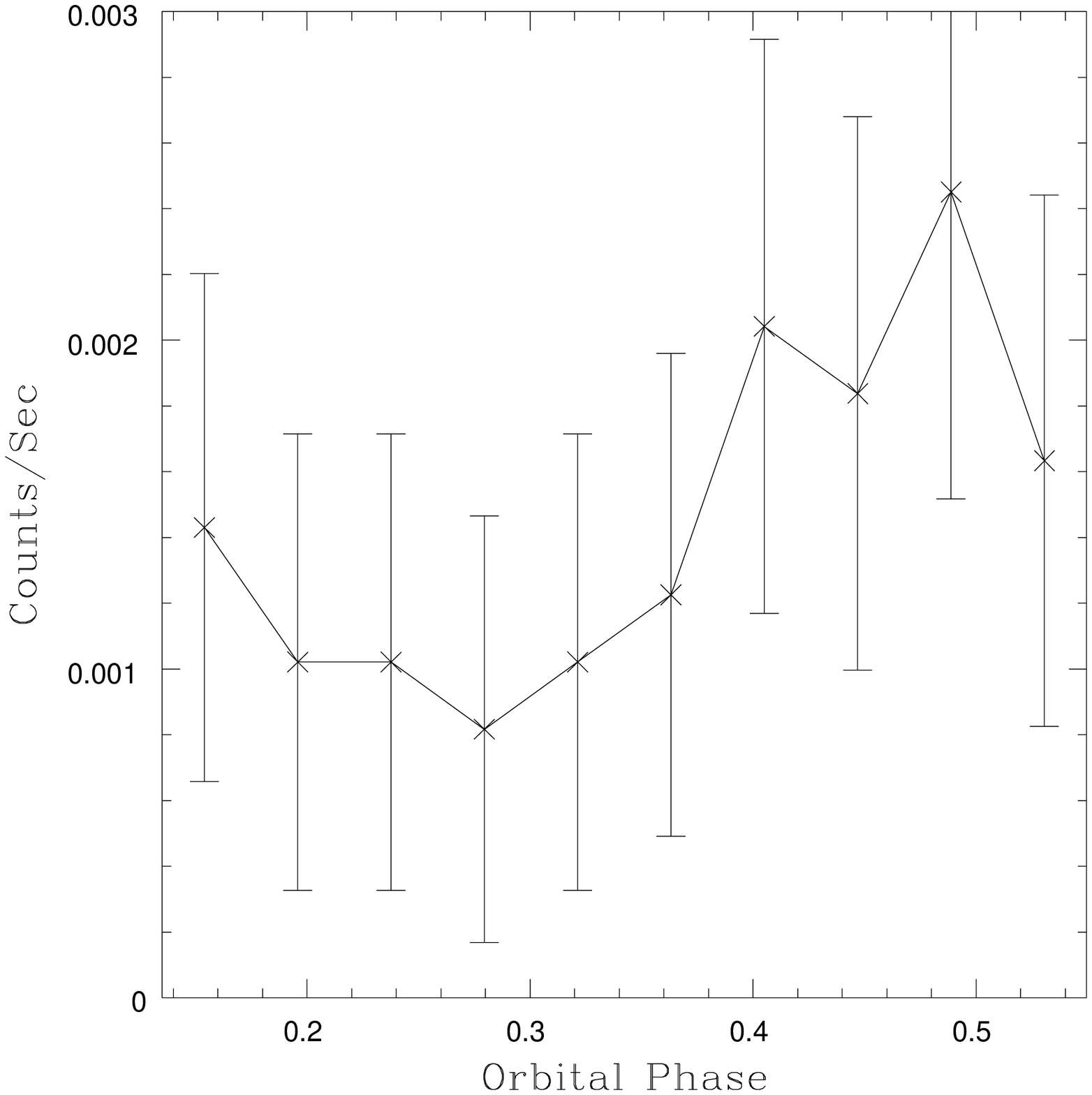,width=2.7in}
\caption{\label{fig:6397} X-ray 
count rate (0.5--4.5\,keV) vs. orbital phase (from
ephemeris of DPM01) for the MSP in NGC 6397, source U12 of GHE01b, observed 
for 49ksec beginning at UT 31.647 July, 2000. 
The source is persistently
eclipsed in the radio at phases 0.05--0.45. The x-ray detection 
at this phase, and increase at phase \gsim0.4, imply at least 
part of the x-ray emission is extended. }
\vspace*{0.14in}
\end{figurehere}

Finally, we note that the large radial offset of 6397-A in the cluster,
at radius \about11\,$r_c$, is much larger than any of the detected, or
plausible new candidate, MSPs in 47~Tuc.  DPM01 and GHE01b noted 
it was likely ejected from the core, 
perhaps in a second exchange encounter since the MSP companion is 
not the expected He WD or low mass (ablated?) remnant of a 
secondary in an LMXB 
(as for all previously identified MSP companions) but rather 
is apparently just off the main sequence. GHE01b noted that 
6397-A is approximately aligned with the central ``line'' of blue
stragglers as well as several other of the Chandra sources ---
suggesting possible asymmetric ejection of binaries from the cluster
core (possibly induced by cluster rotation; R. Spurzem, private 
communication). Since NGC~6397 is a post core collapse cluster with 
a power law cusp measured out to \about60\arcsec (Lugger, Cohn and 
Grindlay 1995) that is 
expected to contain scattered binaries (and NSs), the MSP has surely 
been scattered. 

Burderi et al. (2002) 
show that 6397-A could have evolved from an initial \about14\,h binary 
period to its present 34\,h period, near the critical value where 
pressure from the MSP wind exceeds that of matter overflowing 
the Roche lobe (though intermittent accretion could resume; cf. 
\S 5). However the scattering encounter probably occurred within the 
past \about10$^{7}$ yr since the central relaxation time 
is only \about10$^5$ yr (Harris 1996) and still 
\lsim10$^{6-7}$ yr at the current offset position. 
An upper limit to the scattering time is \about10$^{8-9}$y, the 
likely time since core collapse and also the likely 
MSP ``age'' (Figure 4a). The scattering timescale 
is thus shorter than the binary or nuclear evolution timescales 
leading to the present system, and the  
interaction must have altered the binary parameters. Most probably, 
the binary hardened in the encounter if the present secondary was 
the original donor. This makes it less likely that the secondary, 
now just inside its Roche lobe,  
could have filled its Roche lobe prior to the encounter and spun the 
MSP up in an LMXB phase. If mass transfer occurred only after the 
hardening induced by the encounter, there would not be enough time 
to spin up the MSP given the \mdot \about10$^{-10}$ \Msun yr$^{-1}$ 
for the likely intermittent LMXB phases 
(Burderi et al. 2002) and the MSP ``age''. 

If, however, the current secondary was 
exchanged into the binary displacing what was the original donor 
(probably a much lower mass HeWD, as in most other MSPs), there would not 
be the MSP formation-timescale problem. Even more interesting, this 
may help explain a puzzle pointed out by Orosz \& van Kerkwijk (2002) 
after this paper was submitted: the measured (DPM01) \Edot for 6397-A 
would suggest significant photometric heating of the 
binary companion (by the MSP wind) though no visible heating 
modulation (only ellipsoidal variation) is observed. 
The ``excess'' \Edot might be a factor of 2--3, depending on 
uncertain heat transport in the secondary. 
Orosz \& van Kerkwijk suggest the large \Edot may be due to a \Pdot 
value dominated by a triple companion (easily tested by longer term 
MSP timing). If, instead, the present secondary has been exchanged 
for the original, and it is marginally Roche lobe filling (Burderi 
et al. 2002), then accretion (and LMXB) episodes could lead to spin{\it down} 
episodes of the MSP -- at least for those encounters that are 
retrograde but possibly also for the very misaligned prograde case. 
Depending on uncertain crust-core coupling timescales, 
an increased \Pdot (negative; or positive for the nearly aligned 
prograde case) may persist for the MSP 
between intermittent accretion episodes. This 
enhanced \Pdot in an ``old" MSP with low B field would not produce 
the same enhanced pulsar wind (and thus heating) as a nominal ``young" 
(higher B field) MSP since \Pdot is now dominated by 
internal, rather than electromagnetic, 
rotation loss. The true age of 6397-A would thus be 
correspondingly older. Of course reduction of \Edot would move 6397-A 
off the apparent correlation lines (e.g. Figs. 3, 4a) defined by the 
47~Tuc MSPs, though if (only) a factor of \about2--3 reduction in \Edot 
this might be compensated if only the thermal component of \Lx (not yet 
isolated) were plotted. 

A second example of a possible ``double-exchange" MSP, with a probable  
main sequence secondary, has been discovered with 
the optical identification of the MSP 47~Tuc-W: Edmonds et al. (2002) 
report (after this paper was submitted) the identification of  
Chandra source W29 (cf. GHE01a) with a V=22.3 star in the central core 
(\lsim0.2r$_{c}$ from the center) of 47~Tuc with optical 
modulation identical in both period (to within 0.5\,s) and phase (to 
within 1.2\,min) with the 3.2\,h binary modulation of 47~Tuc-W. The x-ray 
luminosity and hardness ratios for the Chandra source (W29) are 
very similar to those for 6397-A. Although this is a certain MSP 
identification, its one-time (\about4h) radio detection 
(Camilo et al. 2000) precludes measurement of 
its \Pdot and thus \Edot value (possibly enhanced?) so 
that it cannot be included (now) in our analysis of the full 
radio MSP sample in 47~Tuc. Its identification does show, however, 
that our estimates for the total MSP sample in 47~Tuc as derived 
from the soft source distribution are likely to be lower limits.  
Indeed, Edmonds et al. (2002) also find in the core of 47~Tuc 
a second possible MSP with main sequence companion
as the optical counterpart of 
Chandra source W34, with orbital period 97.45m (not matched to 
any of the radio MSP sample). 

\section{DISCUSSION} 
One of the key results of this study is that MSPs in the globular clusters 
47~Tuc 
and NGC~6397 may have a less efficient conversion of
rotational spindown energy ($\dot E$) into soft x-rays (L$_x$) than
most field MSPs, even those with correspondingly low values 
of their magnetic field at the light cylinder. 
For \Lx $\propto$ \Edot$^{\beta}$, the 47~Tuc--NGC~6397 samples are fit by 
$\beta
= 0.5\pm0.15$ whereas the field and M28 sample are consistent with the
value $\beta = 1$ found for non-recycled pulsars. The 4 field MSPs 
with largest \Edot and $B_{\rm lc}$ (cf. Figs. 3 and 5) show 
correlations with \Lx which extrapolate to the low \Edot and $B_{\rm lc}$ 
sample, whereas the correlation shown by the 47~Tuc--NGC~6397 MSPs does 
not extrapolate to higher \Edot and $B_{\rm lc}$ systems. It is possible 
that the difference is just due to the 4 MSPs  with $B_{\rm lc}$ \gsim 
10$^{5.5}$ G being dominated by non-thermal beamed emission 
(e.g. since B1821$-$24 has no binary companion, its
hard spectrum is not from a MSP wind interacting with a
companion, as plausibly is the case for 47~Tuc-J and 6397-A).  
However, then the large deviations from the 47~Tuc correlation of the field 
MSPs at lower $B_{\rm lc}$ must be explained as systematic errors in 
distance, which is unlikely for at least two of them. Perhaps more  
convincing, the 
\Lx/\Edot vs. $\tau$ relations 
for the field vs. 47~Tuc MSPs (Fig. 4b) show significant deviations 
for the low $B_{\rm lc}$ systems. 
What could be different about the MSPs in these two globulars vs. field MSPs? 

The polar cap (PC) heating models of HM02 can predict the 
observed \Lx--\Edot$^\beta$ relation, with $\beta$ \about 0.5, 
for MSPs with spindown ages $\tau = P/2 \dot P$ \gsim 10$^8$ yr that are 
no longer able to produce pairs by curvature 
radiation (CR) $\gamma$-rays (which convert to pairs 
in the pulsar magnetic field) but instead by $\gamma$'s from inverse 
Compton scattering (ICS) of thermal photons from the NS surface. 
>From the numerical models for PC heating by returning positrons, 
with total heating luminosity L$_+$   
for the ICS regime and the resulting prediction for 
L$_+$/\Edot vs. $\tau$ (HM02, Fig. 8b), we 
find the approximate empirical relations   
L$_+$/\Edot $\propto \tau^{\alpha} P^{2\alpha}$ which yields 
L$_{+} \propto$ \Edot$^{1-\alpha}$  and 
L$_{+} \propto \tau^{\alpha - 1}/P^{2(1-\alpha)}$ using 
$\tau \propto$ 1/\Edot $P^2$. 
For MSPs with $\tau$ \lsim 1--2 \X 10$^9$ yr 
(beyond which the expected L$_+$/\Edot decreases as the MSP 
approaches the pair deathline), this gives $\alpha$ \about 0.4--0.5, 
or L$_+$ $\propto$ \Edot$^{\beta}$, or $\beta = 1 - \alpha$ 
= 0.5--0.6. The model predicts (for $P = 5$\,ms and PC temperature 
T = 3 \X 10$^6$ K) log(\Lx/\Edot) \about $-4.5$ + 0.4log($\tau$/10$^9$\,yr)  
and so is consistent with the approximate trend from 6397-A to the 
apparently youngest 47~Tuc MSPs.
However, several key differences 
are evident: (i) the 47~Tuc MSPs (and 6397-A) show an unbroken 
power law dependence on age, with \Lx $\propto L_+ \propto \tau^{-0.3}$ 
(Fig. 4a), well beyond the expected steepening (HM02, Fig. 9b) 
at $\tau$ \about10$^9$ yr, (ii) the \Lx/\Edot relation (Fig. 4b) 
likewise continues to increase beyond the expected flattening 
at similar $\tau \sim 10^9$yr (HM02, 
Fig. 8b). In contrast, the field MSPs 
in Figures 4a,b are more consistent with the HM02 
models, having negative or flat slopes, respectively, at 
$\tau$ \gsim 10$^9$\,yr. However, 
the factor of \about3--10\X excess of \Lx/\Edot over those for 
the 47~Tuc MSPs and the ICS models also suggests these systems 
have an additional source of \Lx: probably non-thermal emission. 
It is interesting  that accounting 
in Figure 4b for the approximate dependence on pulsar period, 
\Lx/\Edot $\propto P^{2\alpha}$ (derived from a fit to 
Fig. 8b of HM02), 
makes the field MSPs have a (slightly) more negative slope 
($-0.16 \pm0.23$ vs. $-0.06 \pm0.21$), as expected from 
HM02 for their age, whereas the 47~Tuc MSPs have a slightly more 
positive logarithmic slope (0.59$\pm$0.10 vs. 0.53$\pm$0.13).   
Finally, (iii) the  
small effective BB emission areas 
(r$_{x}$ \lsim 0.14--0.56km 
for dipole polar caps; smaller still (each) for multipole caps) 
are significantly smaller than the effective radius of the 
annular polar cap emission region predicted (HM02 Fig. 7). 

We suggest that these differences may point to altered surface 
magnetic field configurations for the 47~Tuc (and 6397-A) MSPs 
to enable such efficient PC heating at large age. If their surface 
magnetic fields have 
significant multipole components, the effective B-field 
curvature is greater and the pair production front moves 
closer to the NS yielding more efficient positron return 
to heat the polar caps. The magnetic field footprint on the NS 
becomes smaller, 
giving the small emission areas. 
If the non-thermal radio emission is produced, 
as in most models, from near the light cylinder where the 
field is dominated by the dipole component, then its 
properties are not significantly altered and the radio pulse 
profiles are otherwise similar to field MSPs (as observed; 
Camilo et al. 2000). Clearly this multipole field hypothesis must 
be tested with models (only dipole fields have been considered 
for pulsar emission and PC heating --- e.g. HM02). 
 
Another possibility is that the 
NSs in the 47~Tuc MSPs (and 6397-A) are systematically more 
massive or compact than those in the field MSPs. The models of Harding,  
Muslimov, \& Zhang (2002) show that the ICS 
regime becomes longer lived (extends to lower \Pdot values) 
for more massive or more compact NSs. 
Although our radial 
distribution results (\S 3.2) imply MSP masses (total) 
in the range \about1.1--1.4 \Msun, suggesting the more compact NS 
interpretation may be preferred, more massive NSs are allowed 
if in fact the MSP distribution is ``puffed up'' by 
binary scattering. 

The hard
spectra in a few of these ``thermal'' MSPs (47~Tuc-J and 6397-A) must
have a different origin. In 6397-A, the possible modulation phased 
with the radio eclipse of the hard source suggests the 
hard source is due to shocked gas lifted from 
the binary companion by the MSP wind pressure (Burderi et al. 2002). 
If the hard
spectrum is a power law (thermal bremsstrahlung is ruled out by the 
$\delta$DM constraint [\S 2.1] unless the emission is 
strongly aspherical), 
and the shocked wind density and thus emission 
measure is proportional to the underlying MSP thermal luminosity, then the
approximate agreement of these hard spectrum, but low L$_x$, systems
with the $\beta$ = 0.5 law exhibited by the soft (thermal) MSPs would imply
scaling of the shocked gas emission with the thermal (NS) emission. 
As mentioned above, this might be expected if both the MSP wind 
and PC heating are proportional to the PC current.

What is special about globular cluster MSPs to drive them 
towards altered $B_{\rm surf}$ 
(and thus $B_{\rm lc}$)  values or configurations 
or possibly larger NS mass (or compactness)? 
Unlike MSPs in
the field, those in dense cluster cores have a possibility of being
driven back into contact and an accretion phase, as a re-cycled LMXB
(from a MSP)!~ Renewed accretion (and the 6397-A 
companion is presently close to
filling its Roche lobe; cf.\ DPM01) would likely continue the B-field
burial process thought to be responsible (e.g.\ Romani 1990) for field
decay from the $\ga10^{11}$\,G fields at NS birth to the $\la10^9$\,G
values typical of MSPs. 
This could lead, in turn, to altered B-field configuration 
(multipole field components), particularly for a 
secondary re-exchanged into an MSP (with random spin-orbital 
encounter angular momentum). It would also systematically increase the NS 
mass beyond that for field MSPs. 
 The 47~Tuc MSPs have spent their lifetime $\tau
\ga 1$\,Gyr in a dense ($n \sim 10^5\,{\rm pc}^{-3}$) cluster core,
where they undergo scattering interactions with both single stars and
other binaries.  Such scattering causes the binding energy $x$ of a
hard binary to secularly increase (``hard binaries harden'') at a rate
$\dot{x} \sim 4 m \sigma^2\,\,n {p_0}\!^2 \sigma$ (Goodman \& Hut
1993), where $m$ is the single star mass, $\sigma$ is the
one-dimensional velocity dispersion, and $p_0 = G m/\sigma^2$ is the
impact parameter for a 90\arcdeg\ deflection ($\sim 10$\,AU in the core
of 47~Tuc).  Thus, the binding energy of a typical hard binary increases
by of order $m \sigma^2$ per core relaxation time ($\sim 10^8$\,yr for
47~Tuc).  Over the $\ga 1$\,Gyr lifetime of the MSP binary system, the
total relative binding energy change is $\dot{x} \tau/x \sim 1\%$.
Thus, while angular momentum transfer to the secondary acts to detach
the binary, scattering events tend to drive it back toward contact.

A complication in this picture is that since MSPs are extremely hard
binaries ($x \sim 1000\, m \sigma^2$), $x$ is adiabatically invariant
in weak scattering events where the distance of closest approach much
exceeds the binary separation (e.g.\ Heggie \& Hut 1993).  Thus, the
secular increase in $x$ is largely the result of infrequent strong
scattering events, which are likely to result in an ejection of the
binary from the cluster core 
(e.g. 6397-A) 
or even the cluster.  Other unresolved
issues include the effects of binary-binary scattering, which probably 
``bump'' the MSPs back into contact more effectively than does
binary-single scattering, and orbital eccentricity evolution during
scattering events, which will also tend to drive some binaries towards
contact.  Nevertheless, it appears plausible that a typical old ($\tau
\ga 1$\,Gyr) MSP in a \emph{dense} cluster core might undergo one or
more MSP--LMXB recurrence cycles. Thus a few percent of the MSPs at any
one time may be in (or near) this recurrent LMXB phase. This might
explain the puzzling luminous qLMXBs X5 and X7 (GHE01a and Heinke et
al.~2002) in 47~Tuc:  they could be recently ($\sim10^{5-6}$\,yr)
revived qLMXBs, in which their underlying MSP nature is hidden by 
a heated (by the MSP wind) outflow from the disk and Roche lobe 
of the probable main sequence secondaries (at least for X5, 
which may then be similar to 6397-A and 47~Tuc-W; cf. \S4). 

The (possibly) younger system in NGC~6397 has the advantages of having been
recently scattered out of a still higher density (\about10$^6$
pc$^{-3}$) core collapsed cluster core  and possibly having exchanged
its companion. Either or both would likely have restored an accretion
phase. Thus the MSP need not be just ``born,'' as suggested by Ferraro et
al. (2001); it
may instead have just been reborn.  In contrast, the MSP in M28 is both
\about10\X\ younger, single and in a lower density core and so is unlikely to
have gone through a renewed accretion phase. This MSP--LMXB recycling 
scenario would imply: (i) a population of MSPs with main sequence 
or red straggler 
secondaries (either primordial or re-exchanged) like 6397-A which may 
be largely eclipsed (in radio) by mass loss and 
dominated by hard shocked gas emission, (ii) a 
strong dependence for this multiply-recycled population to occur primarily 
in clusters with largest 
central mass density $\rho_o$ = n$_o$ m and core radius r$_c$, 
and thus total interaction rate  
($\Gamma \propto \rho_o^{1.5} r_c^2$) 
such as Terzan~5, and (iii) that multiple 
MSP--LMXB phases may be more likely, thus yielding $B_{\rm surf}$ burial and  
altered $B_{\rm lc}$  and suppressed magnetospheric emission,  
in MSPs with main sequence companions. 
MSPs with either main sequence or slightly evolved secondaries 
(e.g. 47~Tuc-W or 6397-A; cf. \S~4)  are also most likely to 
contain re-exchanged secondaries which may have anomalous \Pdot 
values (as might 6397-A) from the braking torque supplied by 
a mis-aligned crust-core spin coupling due to the renewed accretion 
phase(s) from the re-exchanged secondary. Even an original 
secondary, if driven back into contact or if marginally filling 
its Roche lobe, could give 
rise to enhanced \Pdot values: when accretion stops, the crust 
is spinning faster than the core, which then exerts a braking 
torque on the crust (and B-field) of the MSP. The timescale for 
equilibrium to be established is uncertain but likely to 
be relevant for MSPs with companions closest to filling 
their Roche lobes (hence the main sequence or subgiant systems) 
which have most recently accreted.  
These main sequence systems, which likely evolve into 
the short period 
(\about1.5--5.5\,h) group identified by Camilo et al. (2000), 
also have longer lifetimes to undergo binary scattering than 
those spun-up from sub-giant companions and now in the longer period 
(\about0.4--2.4\,d) group with He-WD companions. However this lifetime is 
partly offset by the larger binary cross sections for the He-WD group, so 
both are expected to undergo scattering in dense cores. The 
generally larger orbital eccentricities for He-WD secondary MSPs 
in globulars vs. those in the disk has previously been noted (Phinney 1992) 
as evidence for interaction of these MSPs with cluster stars. 
The large eccentricity of 47~Tuc-H (FCL01) suggests it has undergone 
such an encounter. 

\section{CONCLUSIONS}
In this paper we have examined the x-ray properties of the full MSP
populations in 47~Tuc and NGC~6397 localized via radio timing.  The
large sample of MSPs resolved by Chandra in 47~Tuc, and the single
system in NGC~6397, allow a number of conclusions to be drawn:

\begin{itemize}
\item The MSPs in 47~Tuc and NGC~6397 
are surprisingly uniform in \Lx despite a much
larger variation in $\dot E$. The L$_x$--\Edot correlation appears to
be significantly flatter than for a (smaller) sample of MSPs in the
field as well as the isolated (and young) MSP in M28. The correspondingly 
flatter L$_x$--$B_{\rm lc}$ correlation suggests that the 47~Tuc (and NGC~6397)  
MSPs may have $B_{\rm surf}$ and thus $B_{\rm lc}$ configurations 
systematically different (e.g. multipole components) 
from field MSPs and which prolong thermal and suppress magnetospheric 
x-ray emission. 

\item The MSPs in 47~Tuc are dominated by thermal emission sources,
consistent with BB emission from the pulsar polar caps. At least one
(47~Tuc-J) must also have significant non-thermal emission, which may  
arise from the pulsar wind interacting with the companion (see below).

\item The radial distribution of MSPs in 47~Tuc is consistent with the 
predictions of a multi-mass King model for a distribution of 
MSPs with mass ratio $q \sim 2$ times the typical cluster 
stellar mass in the core, i.e.\ a MSP system mass of $\sim 1.4~M_{\odot}$.

\item From the distribution of Chandra sources with similarly soft
x-ray colors, as well as the total counts detected in Soft and Medium
bands vs. cluster radius, the number of MSP candidates with \Lx \gsim 1
\X 10$^{30}$ \lcgs\ is estimated to be \about35--90, allowing for the
plausible contamination of BY Dra and soft CV systems and
incompleteness of the detection for faint sources near the cluster
center. 
The relative number of MSPs detected in the 
complete soft x-ray vs. radio samples suggests 
the radio emission is beamed with beaming factor \about0.2--0.5 
(times any soft x-ray beaming, which is presumably isotropic).

\item A larger population of still fainter MSPs 
($L_X<10^30$ \lcgs) could be detected in x-rays than in the
radio, given the relatively flat L$_x$--L$_{\rm radio}$ correlation
seen for 47~Tuc MSPs. If the excess of Soft vs. Medium band counts in
the cluster core is due to fainter MSPs, they must have relatively lower
luminosity and lower temperatures than the identified MSPs. 

\item The one (binary) MSP system identified in the core-collapsed
globular NGC~6397 is consistent with the L$_x$--\Edot relation of the 47~Tuc
sample despite its very different spectrum, which must be dominated by
non-thermal or bremsstrahlung emission. 
Its harder x-ray emission must (largely) arise
from an extended source, possibly resulting from interaction of the
relativistic wind from the MSP with mass loss from its 
companion star.
The fact that it is still consistent with  the 
\Lx $\propto$ \Edot$^{1/2}$ relation found for the thermal MSPs in 
47~Tuc suggests that both the pulsar wind luminosity and polar cap heating are 
governed by the total polar cap current ($\propto$ \Edot$^{1/2}$; cf. HM02). 
If \Pdot is enhanced by renewed accretion or second exchanges, the 
electromagnetic spindown rate and corresponding 
reduced \Edot may correlate with the
thermal component of \Lx (to be measured) as for 47~Tuc.

\item The enhanced thermal luminosity, small emission 
areas and extension to larger MSP 
ages, for the 47~Tuc MSPs vs. the HM02 models suggests the surface 
magnetic field might be multipole (dipole at the light cylinder 
radius and probable radio emission region), possibly by field 
perturbations from repeated accretion events due to encounters 
or re-exchanges of the secondaries of these MSPs in dense cluster cores. 
Alternatively, 
the extended pair heating (and thermal x-rays) could arise (Harding,  
Muslimov, \& Zhang~2002) if the 
NSs in these MSPs are more massive or (given their radial 
distribution) more compact.
If re-exchanged secondaries are present (as perhaps for 6397-A 
and 47~Tuc-W), apparent \Pdot and thus \Edot values may be enhanced by 
core-crust coupling drag effects over their purely 
electromagnetic spindown values which 
give rise to heating their companions.  
Some evidence for a similar ``excess'' 
\Edot value is available for 47~Tuc-W (Edmonds et al. 2002), but 
as with 6397-A these estimates are subject to uncertain 
heating efficiencies and transport effects (which might heat the backside 
of the tidally-locked secondary).

\item NGC~6397 appears to be deficient in its total MSP population per
unit mass compared with 47~Tuc (GHE01b), if most cluster MSPs are like
the 47~Tuc sample and dominated by soft thermal sources.  Only 3--5 such
objects are found in the sample of 20 Chandra sources detected, with
\Lx limits a factor of 2 lower than in 47~Tuc (GHE01b). Most of the other
Chandra sources in NGC~6397 with comparable x-ray colors (or actual
spectra) are identified with CVs (optically; cf.  GHE01b and references
therein) and thus not likely to be additional MSPs. However it is
possible that a few of the main sequence binaries, or BY Dra candidates
(TGE01) could also harbor MSP primaries as is the case for 6397-A.

\item Conversely, the very large possible BY Dra population in 47~Tuc
(up to \about10$^3$ are possible; R.~Gilliland, private communication) may
contain a significant sample of MSPs as may have occurred in 6397-A:
those that have likely had a cluster main sequence star exchanged {\em
into\/} the binary, displacing the probable degenerate secondary which had
evolved from the original mass donor to spin up the MSP. 
If the 6397-A system with such a secondary 
(now evolved or bloated; cf. \S~4) is typical, and mass loss is
more readily driven off its near Roche-lobe filling secondary 
by the MSP wind pressure (Burderi et al. 2002), 
then these
systems will likely be dominated by shocked gas and appear
hard. Since these main sequence secondary MSPs  will 
typically have larger MSP heating (given their larger 
secondaries) and thus mass loss, 
they may be largely eclipsed until their secondaries have been ablated to 
become the very low mass group (\about0.02\,$M_\odot$) found (Camilo et al. 
2000) in the 47~Tuc MSPs. Thus the total MSP sample in 47~Tuc 
would include faint hard
sources as well as the soft, thermal-dominated, systems and the total 
MSP population would exceed the \about35-90 estimated from the 
soft source population alone. 

\end{itemize}

We shall obtain a very deep (300ksec) cycle 3 Chandra observation of
47~Tuc to probe the complete MSP population as well as to better
constrain and measure x-ray colors and actual spectra to test emission
models. The hard spectra MSPs (e.g. 47~Tuc-J) may be variable (as is
6397-A), and hard components should be detectable in other eclipsing
MSPs (e.g. 47~Tuc-O).  The interlocking puzzles of compact binary
populations (e.g.  MSPs vs. BY Dra systems) and compact object
populations (MSPs vs. CVs, as measures of NS vs. WD populations) in
globular clusters offer new insights into the stellar and dynamical
evolution of these oldest stellar systems in the Galaxy.

\acknowledgments

We thank Paulo Freire and Michael Kramer for help with and discussions
regarding the radio data, Ron Gilliland for help with the 
HST data, Werner Becker for comments on the paper, Alice Harding 
for comments on PC models, Fred Lamb for comments on core-crust 
coupling effects and 
the referee Dick Manchester for suggestions.  
This work was supported in part by NASA
grants GO0-1098A and HST-AR-09199.01-A (JG) and SAO grant 
GO1-2063X (FC).

\clearpage

\begin{deluxetable}{lccccccr}
\tablecaption{\label{tab:results} Chandra Results on MSPs in 47~Tuc and
NGC~6397 }
\footnotesize
\tablewidth{0pt}\tablehead{
\colhead{MSP}           &
\colhead{$\Delta$RA\tablenotemark{a}}    &
\colhead{$\Delta$Dec\tablenotemark{a}}   &
\colhead{Softcts\tablenotemark{b}}       &
\colhead{Mediumcts\tablenotemark{b}}     &
\colhead{Hardcts\tablenotemark{b}}       &
\colhead{log(L$_x$)\tablenotemark{c}}    &
\colhead{log($\dot E$)\tablenotemark{d}} \\
 & (arcsec) & (arcsec) & (0.2--1\,keV) & (1--2\,keV) & (2--8\,keV) & 
(0.5--2.5\,keV) & \\
}
\startdata
47~Tuc-C & \nodata  & \nodata &     1 &  0 &  0    & 29.6 & 33.2\\
47~Tuc-D & $-$0.17  &    0.19 &     5 &  7 &  1    & 30.3 & 33.7 \\
47~Tuc-E &    0.01  & $-$0.07 &    11 & 13 &  0    & 30.6 & 34.5 \\
47~Tuc-F &    0.32  &    0.06 &    10 &  9 &  1    & 30.5 & 34.4 \\
47~Tuc-G &    0.05  &    0.03 &     4 &  4 &  0    & 30.1 & 33.7 \\
47~Tuc-H &    0.03  &    0.01 &     5 &  4 &  0    & 30.1 & 33.2 \\
47~Tuc-I &    0.16  &    0.01 &     5 &  5 &  1    & 30.2 & 33.9 \\
47~Tuc-J & $-$0.03  & $-$0.07 &     3 &  5 &  5    & 30.3 & 34.4 \\
47~Tuc-L & \nodata  & \nodata & 12(10)&  5 &  6(0) & 30.4 & 34.2 \\
47~Tuc-M &    0.36  & $-$0.01 &     2 &  3 &  0    & 30.1 & 33.5 \\
47~Tuc-N &    0.01  &    0.10 &     5 &  6 &  0    & 30.2 & 34.1 \\
47~Tuc-O &    0.10  &    0.19 &    17 & 11 &  0    & 30.6 & 34.4 \\
47~Tuc-Q & \nodata  & \nodata &     5 &  3 &  0    & 30.1 & 34.3 \\
47~Tuc-S & $-$0.26  & $-$0.39 &     5 &  4 &  1    & 30.2 & \nodata \\
47~Tuc-T & \nodata  & \nodata &     4 &  3 &  0    & 30.0 & 33.8 \\
47~Tuc-U &    0.04  & $-$0.06 &     4 &  7 &  1    & 30.3 & 34.2 \\
6397-A  & $-$0.05  &    0.09 &    11 & 40 & 20    & 30.9 & 35.1 \\
\enddata
\tablenotetext{a}{$\Delta$RA and $\Delta$Dec offsets are Chandra
\WAVDETECT minus precise MSP radio (timing) position (47~Tuc) and
Chandra \WAVDETECT minus HST position (NGC~6397). 
The close pair  
47~Tuc-F, -S are marginally resolved (see text) and thus have larger 
offsets; offsets are not listed for -L, -Q, -T and -C, without \WAVDETECT 
positions due to crowding or limited counts (see Fig. 1).}
\tablenotetext{b}{Counts in each band derived from 1\arcsec aperture on
radio MSP position.  MSPs 47~Tuc-G and -I are indistinguishable; counts
are tentatively divided  based on respective $\dot E$ values.  MSPs
47~Tuc-F and -S detected combined with {\sc wavdetect}; counts estimated
separately (see text).  47~Tuc-L suffers severe crowding; values in
parentheses indicate best estimates removing other source
contamination. }
\tablenotetext{c}{\Lx values (in \lcgs) are computed (PIMMS) for an
assumed blackbody spectrum with kT=0.22\,keV (see text) and include
exposure correction factors of 2.5, 1.4 and 1.9 for 47~Tuc-C, -J, and
-M, respectively (all other MSPs detected with uniform exposure factor
1.0). Approximate errors on log L$_x$ are $\pm0.3$ for 47~Tuc-C and
$\pm0.2$ for the others. }
\tablenotetext{d}{\Edot values (in \lcgs) are derived for the 47~Tuc
MSPs assuming the MSPs are at 3D radii appropriate to their $\delta$DM,
determined using a constant gas density in the cluster (Freire et al.~2001b), 
and a
King model for the cluster potential with central velocity dispersion
11.6\,km\,s$^{-1}$. \Edot value for 47~Tuc-S is indeterminate. \Edot for
the MSP 6397-A, at relatively large radial offset in NGC~6397, is
derived assuming the cluster acceleration contribution is small.}
\end{deluxetable}


\begin{deluxetable}{lcccccl}
\tablecaption{\label{tab:mspdata} Parameters for MSPs Previously
Detected in X-rays }
\footnotesize
\tablewidth{0pt}\tablehead{
\colhead{MSP}        &
\colhead{$P$}        &
\colhead{$\dot P_i$} &
\colhead{$d$}        &
\colhead{$S_{1400}$} &
\colhead{$F_x$\tablenotemark{a}} &
\colhead{References} \\
                     & 
                (ms) &
  $(\times10^{-20})$ &
               (kpc) &
               (mJy) &
 $(\times10^{-13}$\,erg\,s$^{-1}$\,cm$^{-2})$ &
                     \\
}
\startdata
J0030$+$0451 &3.05 &$<$1.0  &   0.23 &   0.6  &  2.5  & 1, 2     \\ 
J0218$+$4232 &2.32 &   7.78 &$>$5.9  &   0.9  &  1.5  & 3, 4     \\
J0437$-$4715 &5.76 &   1.86 &   0.14\tablenotemark{b} & 137   & 19. & 4, 5, 6\\
J0751$+$1807 &3.48 &   0.80 &   2.02 &   1.0  &  0.83 & 4, 7     \\
J1012$+$5307 &5.26 &   0.97 &   0.84 &   2.8  &  0.49 & 4, 7, 8  \\
J1024$-$0719 &5.16 &$<$0.30 &$<$0.20\tablenotemark{b} &  0.9  & 0.2 & 4, 6, 9\\
J1744$-$1134 &4.07 &   0.71 &   0.36\tablenotemark{b} &  2.0  & 0.2 & 4, 6, 
10\\
B1937$+$21   &1.56 &  10.6  &$>$3.6\tablenotemark{b}  &  16  & 3.7 & 7, 9, 11\\
B1957$+$20   &1.61 &   1.15 &   1.53 &   0.35 &  3.0  & 4, 7, 9  \\
J2124$-$3358 &4.93 &   1.30 &   0.25 &   2.6  &  2.5  & 4, 6, 9  \\
B1821$-$24 (M28) &3.05 & 162.   &   5.5\tablenotemark{b}  & 1.9  & 5.2 & 4, 7 
\\
\enddata
\tablerefs{1 (Lommen et al.~2000); 2 (Becker et al.~2000); 3 (Navarro
et al.~1995); 4 (BT99); 5 (van Straten et al.~2001); 6 (Toscano et
al.~1998); 7 (Taylor et al.~1995); 8 (Lange et al.~2001); 9 (Toscano et
al.~1999b); 10 (Toscano et al.~1999a); 11 (Takahashi et al.~2001). }
\tablenotetext{a}{X-ray flux in ROSAT (0.1--2.4\,keV) band.  Value for
B1937+21 in ASCA (2--10\,keV) band. }
\tablenotetext{b}{Accurate distances/limits.  Other estimates are
obtained primarily from dispersion measures together with the Taylor \&
Cordes (1993) electron density model, and are rather uncertain. }
\end{deluxetable}


\begin{deluxetable}{lcccc}
\tablecaption{\label{tab:ks} Results from KS Two-Sample Comparisons }
\footnotesize
\tablewidth{0pt}
\tablehead{
\colhead{} &
\colhead{MSP} &
\colhead{\parbox[t]{0.6in}{\centering{Medium\\(raw)\\*[-6pt]~~}}} &
\colhead{\parbox[t]{0.6in}{\centering{Medium\\(corr)\\*[-6pt]~~}}} &
}
\startdata
Soft           &  0.66  &  0.0034   &  0.11     \\
Medium (raw)   &  0.16  &  \nodata  &  \nodata  \\
Medium (corr)  &  0.82  &  \nodata  &  \nodata  \\
\enddata
\tablecomments{The entries give the probability of the null hypothesis
that the two samples are drawn from the same underlying distribution.
Values of less than 0.05 indicate significant differences between the
samples. }
\end{deluxetable}


\begin{deluxetable}{lcrcrc}
\tablecaption{\label{tab:king} Results from Generalized King Model Fits
}
\footnotesize
\tablewidth{0pt}
\tablehead{
\colhead{Sample \tablenotemark{a}} & 
\colhead{$N$} &
\colhead{$r_c~('')$\tablenotemark{b}} &
\colhead{$\alpha$}
}
\startdata
Soft                   & 44 &  $15.2 \pm \phn3.5$ &  $-3.6 \pm 0.8$ \\
Medium (raw)           & 41 &  $ 5.6 \pm \phn3.9$ &  $-1.9 \pm 0.3$ \\
Medium (corr)          & 33 &  $ 9.0 \pm \phn6.9$ &  $-2.5 \pm 1.0$ \\
MSP                    & 16 &  $13.1 \pm \phn6.9$ &  $-2.4 \pm 1.1$ \\
CV                     & 19 &  $10.3 \pm    10.0$ &  $-3.1 \pm 3.7$ \\
BY + RS \tablenotemark{c}                & 14 &  $15.8 \pm \phn4.5$ &  $-5.6 
\pm 
4.1$ \\
No ID                  & 22 &  $17.5 \pm \phn3.6$ &  $-3.1 \pm 0.7$ \\
\enddata
\tablenotetext{a}{Samples were delimited by a radial cutoff of
250\arcsec except for the MSP group for which 100\arcsec was used. }
\tablenotetext{b}{Core radius for x-ray source population 
(vs. {$r_c~$} = 24.0\arcsec for the optical; see text).}  
\tablenotetext{c}{BY + RS sample is the combination of BY Dra (main 
sequence) binaries and some possible RS CVn (sub-giant) systems.}
\end{deluxetable}


\begin{deluxetable}{llcc}
\tablecaption{\label{tab:cands} Candidate MSPs from Soft Sources }
\footnotesize
\tablewidth{0pt}\tablehead{
\colhead{RA\tablenotemark{a}} & 
\colhead{Dec\tablenotemark{a}} &
\colhead{Medcts} &
\colhead{log(L$_x$)\tablenotemark{b}}\\
 \colhead{(J2000)} & \colhead{(J2000)} & \colhead{(0.5--4.5\,keV)} & 
 \colhead{(0.5--2.5\,keV)} \\
}
\startdata
00:24:13.45(1) & $-$72:04:51.27(8) & 33.5 & 30.7 \\
00:24:12.53(1) & $-$72:04:41.13(5) &  8.3 & 30.1 \\
00:24:11.64(3) & $-$72:05:03.72(8) &  9.1 & 30.1 \\
00:24:07.38(2) & $-$72:04:49.41(5) & 41.8 & 30.8 \\
00:24:06.05(2) & $-$72:05:01.7(1)  & 12.7 & 30.3 \\
00:24:04.64(2) & $-$72:04:34.5(1)  & 18.7 & 30.5 \\
00:24:03.22(2) & $-$72:04:29.74(8) & 15.7 & 30.4 \\
00:24:03.14(2) & $-$72:04:26.55(9) & 16.7 & 30.4 \\
00:24:02.63(3) & $-$72:05:38.5(1)  & 13.8 & 30.3 \\
00:24:02.52(2) & $-$72:04:44.1(1)  & 20.4 & 30.5 \\
00:23:59.91(2) & $-$72:05:04.7(1)  & 21.2 & 30.5 \\
00:23:59.45(2) & $-$72:04:47.9(2)  &  9.4 & 30.2 \\
00:24:01.45(3) & $-$72:04:41.6(1)  &  9.7 & 30.2 \\
00:24:01.04(2) & $-$72:04:13.9(1)  & 12.5 & 30.3 \\
00:23:59.45(3) & $-$72:04:38.5(1)  &  9.9 & 30.2 \\
00:24:05.03(4) & $-$72:05:02.8(1)  &  9.0 & 30.1 \\
00:24:17.20(2) & $-$72:03:03.2(1)  &  8.7 & 30.1 \\
00:24:11.50(1) & $-$72:02:24.7(1)  & 20.6 & 30.5 \\
00:24:42.49(4) & $-$72:06:20.6(2)  & 15.1 & 30.4 \\
\enddata
\tablecomments{The first 5 source entries, and the 4th from 
the bottom of the table, are all particularly promising MSP 
candidates as they are all included in the Albrow et al. (2001) 
HST field and yet lack evidence for optical counterparts with 
either CV or BY Dra characteristics.}
\tablenotetext{a}{The RA and Dec have been corrected to the MSP
coordinate frame. }
\tablenotetext{b}{\Lx values calculated as for
Table~\ref{tab:results}.  }
\end{deluxetable}

\end{document}